\documentclass[a4paper,twocolumn,11pt, unpublished]{quantumarticle}
\pdfoutput=1

\usepackage{xcolor}
\definecolor{Cornflower}{RGB}{100, 149, 237}

\usepackage[normalem]{ulem}  

\usepackage{qcircuit}
\usepackage{array}
\usepackage{textcomp}
\usepackage{physics}
\usepackage{float}
\usepackage{graphicx}
\usepackage{amsmath,mathtools}
\usepackage{amssymb}
\usepackage[colorlinks=true,
            citecolor=blue,
            linkcolor=blue,
            urlcolor=blue,
            filecolor=blue,
            pdfborder={0 0 0}]{hyperref}
\usepackage{setspace}
\usepackage{indentfirst}
\usepackage[toc, page]{appendix}
\usepackage{CJKutf8}
\usepackage[numbers,sort&compress]{natbib}

\usepackage[utf8]{inputenc}
\usepackage[english]{babel}
\usepackage[T1]{fontenc}
\usepackage{amsmath}
\usepackage{hyperref}

\def\a{\alpha}
\def\b{\beta}

\def\d{\delta}
\def\D{\Delta}

\def\g{\gamma}

\def\k{\kappa}

\def\r{\rho}
\def\s{\sigma}

\newcommand{\bE}{{\mathbb E}}

\newcommand{\bI}{{\mathbb I}}

\newcommand{\cO}{{\mathcal O}}

\def\Expval#1{\bE\qty[#1]}
\def\dress#1{\widetilde{#1}}

\def\Var#1{\operatorname{Var}\qty[#1]}

\begin{document}


\title{Scrambling Dynamics with Imperfections in a Solvable Model}

\author{Nadie LiTenn(\begin{CJK*}{UTF8}{gbsn}李依落\end{CJK*})}
\affiliation{Department of Physics, Brandeis University, Waltham, Massachusetts USA}

\author{Tianci Zhou}
\affiliation{Department of Physics, Virginia Tech, Blacksburg, Virginia 24061, USA}

\author{Brian Swingle}
\affiliation{Department of Physics, Brandeis University, Waltham, Massachusetts USA}

\email{bswingle@brandeis.edu}

\date{June 2, 2025}

\begin{abstract}

We study how probes of quantum scrambling dynamics respond to two kinds of imperfections --- unequal forward and backward evolutions and decoherence --- in a solvable Brownian circuit model. We calculate a ``renormalized'' out-of-time-order correlator (ROTOC) in the model with $N$ qubits, and we show that the circuit-averaged ROTOC is controlled by an effective probability distribution in operator weight space which obeys a system of $N$ non-linear equations of motion. These equations can be easily solved numerically for large system sizes which are beyond the reach of exact methods. 
Moreover, for an operator initially concentrated on weight one $w_0=1$, we provide an exact solution to the equations in the thermodynamic limit of many qubits that is valid for all times, all non-vanishing perturbation strengths $p\gtrsim 1/\sqrt{N}$, and all decoherence strengths. We also show that a generic initial condition $w_0 >1$ leads to a metastable state that eventually collapses to the $w_0=1$ case after a lifetime $\sim \log(N/w_0)$. Our results highlight situations where it is still possible to extract the unperturbed chaos exponent even in the presence of imperfections, and we comment on the applications of our results to existing experiments with nuclear spins and to future scrambling experiments.
\end{abstract}


\maketitle


\section{Introduction}
\addtocontents{toc}{\protect\contentsline{section}{}{}{}}

Scrambling is a physical process by which local information is rendered complex and effectively inaccessible to simple probes. It has found relevance in fields including quantum metrology \cite{Li2023a, Kobrin2024, Davis2016, Zhang2025, Goldstein2011, Macri2016}, chaos and black holes \cite{ Shenker2014, Sekino2008, Hayden2007, Maldacena2016a, Xu2020},
and thermalization of quantum mechanical systems \cite{Srednicki1994, Deutsch1991, Murthy2019}. It can be usefully diagnosed using out-of-time-order correlators (OTOCs) \cite{Larkin1969,Shenker2014, Lantagne-Hurtubise2020} and is understood microscopically within a framework of operator growth \cite{Xu2023, Nahum2018, Keyserlingk2018, Hosur2016, Xu2019, Chen2018, Jonay2018a, Kitaev2015, Lewis-Swan2019, Swingle2018a}. There is also significant experimental interest in measuring scrambling, including many proposals and pioneering experiments \cite{Swingle2016,yao2016interferometricapproachprobingfast, Garttner2017,PhysRevX.9.021061, Braumuller2022, Mi2021, PhysRevX.11.021010, Landsman2019a, Joshi2020}. Moreover, related observables have been measured under the guise of ``multiple quantum coherences'' in ensembles of nuclear spins for some years \cite{Alvarez2010,  Alvarez2011, Alvarez2015, Dominguez2021, Dominguez2021a, Wei2018}.

A crucial question raised by the experiments is how to discern signatures of the underlying unitary scrambling dynamics in a system subject to noise and other experimental imperfections. Several early works have addressed this by proposing methods to approximately remove the effects of noise \cite{Swingle2018} or by proposing protocols that isolate the unitary dynamics of interest by post-selection \cite{Yoshida2019}, and related methods have been used in early experimental studies \cite{Alvarez2015, Landsman2019b}. Several works have also studied how open-system effects modify the signatures of scrambling \cite{Haikka2012, Schuster2023, Syzranov2018, Zhang2019}. 

\begin{figure}[t]
    \centering
    \includegraphics[width=0.8\linewidth]{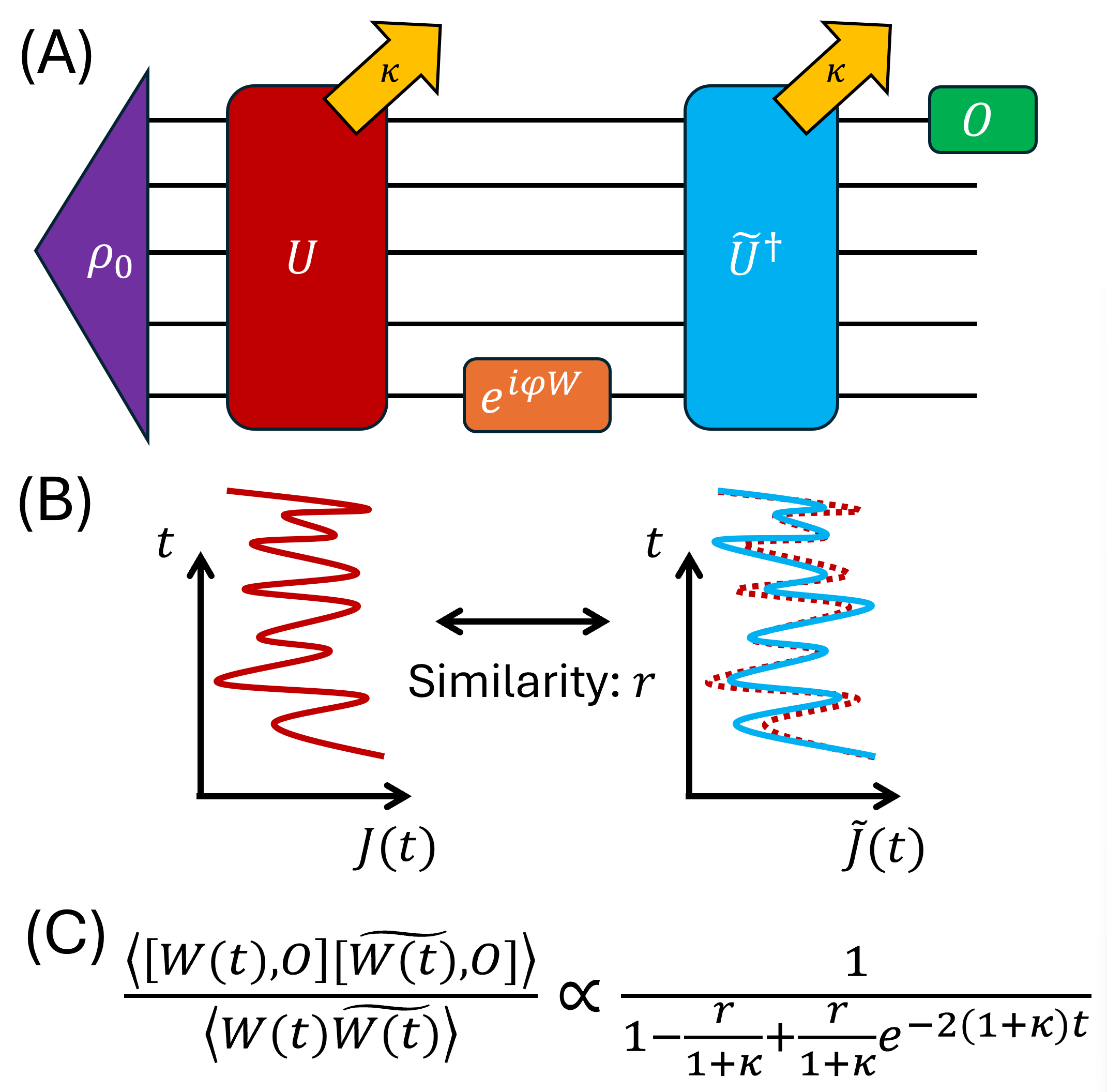}
    \caption{ (A) The protocol to measure the general observable $f_{\rm isolated} (\varphi)$ in Eq.~\eqref{eq:f_isolated} from which we extract the dressed OTOC (numerator of the ROTOC in (C)) and echo (denominator of the ROTOC in (C)). (B) In our Brownian cluster model, the forward (red) and backward (blue) couplings are correlated white noises with normalized covariance $r \in [0,1]$. (C) [Left] The key observable, the ``renormalized OTOC'' or ROTOC, obtained from $f_{\rm isolated}(\varphi)$ ; [Right] our analytic result for the ROTOC in the dilute limit.}
    \label{fig:main_results}
    \vspace{-13pt}
\end{figure}

Beyond simply removing the effects of noise, one can wonder if there might even be new physics in such open and imperfect systems. We were motivated to undertake the present study by just such a possibility: the report of a localization-like phase transition in an ensemble of nuclear spins arising from unequal forward and backward time evolution steps \cite{Alvarez2015}, as shown in Fig.~\ref{fig:main_results}. What was observed in \cite{Alvarez2015} was a change in the behavior of an out-of-time-order correlator from apparent unbounded growth to saturation as the difference between the forward and backward evolutions was increased, thus raising an interesting question about the nature of this transition.

Our goal in this paper is to shed new light on these questions by investigating scrambling with imperfections in a solvable Brownian circuit model. We study a generalized notion of OTOCs which are defined via a protocol that includes unequal forward and backward evolutions and depolarizing noise. This protocol is motivated by experiments with nuclear spins, such as those reported in \cite{Alvarez2015}, but the physics we investigate is more broadly relevant. Using a variety of analytical and numerical results, we are able to understand more deeply the ratio of OTOC with respect to the echo  discussed in \cite{Swingle2018} and are able to microscopically derive quantitative results  that are consistent with \cite{Schuster2023}.

The model we consider is an all-to-all Brownian circuit model~\cite{Lashkari_2013} defined on $N$ qubits with time-dependent Hamiltonian
\begin{equation}
    H(t) = \sum_{i< j,\alpha \beta}^N J_{ij}^{\alpha\beta}(t) \sigma^\alpha_i \sigma^\beta_j = \sum_A J_A O_A,\label{eq:brown_ham}
\end{equation}
which includes all possible 2-body interactions, where $i,j$ label sites, $\a,\b$ label Pauli indices, $A$ is a shorthand for the multi-index, and $O_A\equiv\s_i^\a\s_j^\b$. The coupling constants $J_A(t)$ are independent Gaussian random variables with zero mean and covariance
\begin{equation}
    \mathbb{E}[ J_A(t) J_{A'}(t') ] = \frac{\gamma}{12(N-1)} \delta_{AA'} \delta(t-t'). \label{eq:bc_covar}
\end{equation}
We measure time in units of $\gamma^{-1}$ and hereafter set $\gamma=1$ unless otherwise stated.

We consider two kinds of imperfection, as illustrated in Fig.~\ref{fig:main_results}: i) Hamiltonian perturbations that lead to different forward and backward evolutions ($U \neq \tilde{U}$), and ii) external noise in the form of decoherence ($\kappa > 0$). 

To model perturbations, we will consider distinct couplings for ``forward'' ($J_A$) and ``backward'' ($\dress{J}_A$) evolutions (see Sec.~\ref{sec:methods} and Fig.~\ref{fig:main_results}) with correlation $r$:
\begin{equation}
    \mathbb{E}[J_A(t) \dress{J}_{A'}(t')] = r \mathbb{E}[J_A(t) J_{A'}(t')]. \label{eq:forw_back_covar}
\end{equation}
Here $r=1$ corresponds to identical forward and backward evolutions and $r=0$ corresponds to completely uncorrelated evolutions. We use $\dress{\cdot}$ to denote the imperfection dressed quantities. To model noise, we consider a depolarizing channel that acts with a rate $\kappa$. See Sec.~\ref{sec:methods} and further in App.~\ref{app:protocol} for a detailed discussion of the setup summarized in Fig.~\ref{fig:main_results}.

Our key findings, detailed in Sec.~\ref{sec:results}, reveal a surprising robustness in quantum scrambling at early times and novel features at late times. We derive the microscopic equations governing operator growth in this generalized setup and solve them to obtain the dynamics of a circuit-averaged ``renormalized OTOC'' (ROTOC), shown in Fig.~\ref{fig:main_results}(C). Focusing on the thermodynamic limit, $N \to \infty$, at fixed $r<1$ and $\kappa \geq 0$, which we refer to as the ``dilute limit'', we obtain an exact expression for the renormalized OTOC,
\begin{equation}
    \text{ROTOC } = \frac{8}{3N} \frac{1}{(1-r_{\text{eff}}) + r_{\text{eff}} e^{-2(1+\kappa)t}}
\end{equation}
where $r_{\text{eff}} = \frac{r}{1+\kappa}$ and the time unit is set such that the Lyapunov exponent is $2$ without imperfection. In particular, we find that the system exhibits exponential growth of the ROTOC at early times, with a rate which is unaffected by $r$ and modified in a simple way by $\kappa$. Via the combined parameter $r_{\text{eff}}$, our results also show that imperfect time-reversal and decoherence produce qualitatively similar effects on scrambling dynamics, confirming a conjecture in \cite{Schuster2023}. Moreover, while the result just quoted is for the case where the $W$ operator is a single Pauli, we find that more complex initial conditions consisting of Pauli strings with $w_0$ non-identity elements give rise to a metastable state in which the ROTOC is $w_0$ times the result quoted above. This metastable state has a lifetime of order $\sim\log(N/w_0)$ and eventually decays back to the $w_0=1$ result. Taken together, these results provide a theoretical framework for interpreting experimental measurements of scrambling in the presence of imperfections. As one application, we argue that the signatures of a transition seen in~\cite{Alvarez2015} are likely a finite time crossover effect.


%

The remainder of this paper is organized as follows. In Sec.~\ref{sec:methods}, we define the observables of interest, specify the model, and derive the equations of motion that govern the dynamics. In Sec.~\ref{sec:results}, we discuss the physics arising from the solutions of the equations of motion and its potential application to experiments with nuclear spins. In Sec.~\ref{sec:outlook}, we present an outlook.

\section{The Renormalized OTOC\label{sec:methods}}

\subsection{Observables\label{subsec:observables}}
\label{subsec:methods_obs}

We first define the observables of interest using a protocol that abstracts experiments with nuclear spin ensembles~\cite{Alvarez2010,  Alvarez2011, Alvarez2015, Dominguez2021, Dominguez2021a}. We comment on the potential application of our results to nuclear spins in Sec.~\ref{subsec:results_nmr}. An ideal OTOC not subject to imperfections has the same unitary $U$ for both the forward and backward time evolutions. But our framework can host distinct forward ($U$) and backward ($\dress{U}^\dagger$) evolutions as well as decoherence effects.

Consider $N$ qubits and fix a traceless Hermitian operator $V$, another Hermitian operator $W$, a forward evolution $U$, and a backward evolution $\dress{U}^\dagger$. The following protocol follows the experimental setup to measure the ``dressed'' OTOC $\langle [W(t),V]^\dagger [\dress{W}(t), V]\rangle_\infty $, and ``echo'' $\langle W(t) \dress{W}(t)\rangle_\infty$ where $W(t)=U^\dagger W U$, $\dress{W}(t) = \dress{U}^\dagger W \dress{U}$. Here, $\langle \cdots \rangle_\infty$ denotes the infinite temperature expectation value. We will omit the $\infty$ subscript from here on.

Here, we present the protocol for an isolated system with only perturbation as an imperfection. We can easily modify it to include decoherence by promoting the unitary time evolution to a quantum channel $U\rightarrow D_U (\rho)$ (see App.~\ref{app:protocol} for the full protocol). The protocol is:
\begin{enumerate}
    \item Prepare the initial state, $\rho = \frac{I + \epsilon V}{2^N}$;
    \item Evolve forward, $\rho \to U \rho U^\dagger$;
    \item Rotate, $\rho \to e^{i \phi W}\rho e^{-i \phi W}$;
    \item Evolve backward, $\rho \to \dress{U}^\dagger \rho \dress{U}$;
    \item Measure $V$.
\end{enumerate}
By composing all these steps, the expected value of the final $O$ measurement in the final state $\rho_f$ is
\begin{align}
    \tr(\rho_f V) & = \frac{\epsilon}{2^N}\tr\left(V \dress{U}^\dagger e^{i \phi W} U V U^\dagger e^{-i \phi W} \dress{U}  \right) \nonumber\\
    &= \epsilon \expval{ V \dress{U}^\dagger e^{i \phi W} U V U^\dagger e^{-i \phi W} \dress{U} } \nonumber\\ 
    & = \epsilon f_{\text{isolated}}(\phi,U,\dress{U}). \label{eq:f_isolated}
\end{align}

The last line defines $f_{\text{isolated}}$. From $f_{\text{isolated}}$ we obtain the dressed OTOC $\dress{C}_{W,V}$ via
\begin{equation}
     \dress{C}_{W,V}=\partial_\phi^2 f_{\text{isolated}}(\phi)\big|_{\phi=0} = \expval{[W,\dress{U} V \dress{U}^\dagger][W,U V U^\dagger]},
\end{equation}
and the echo via
\begin{equation}
     f_{\text{isolated}}(\phi)\big|_{\phi=0} = \expval{ \dress{U} V \dress{U}^\dagger U V U^\dagger }.\label{eq:echo_kappa=0}
\end{equation}
When $\dress{U}=U$, the dressed OTOC reduces to the ideal OTOC and the echo is just a constant. For general $U$ and $\dress{U}$, both are non-trivial functions of time. We are primarily interested in the ratio, the ``renormalized OTOC'',
\begin{equation}
   \text{ROTOC} =  \frac{\partial_\phi^2 f}{f}\bigg|_{\phi=0}.
\end{equation}

\subsection{The Brownian Circuit Model \label{subsec:model}}
\label{subsec:methods_model}

Next, we explain how to compute the ROTOC in a Brownian cluster model; for details and a review of the ideal OTOC see App.~\ref{app_master_derivation}. The Brownian cluster model, Eq.~\eqref{eq:brown_ham}, can be defined operationally as the limit of a discrete time evolution with Hamiltonians of varying Gaussian couplings. We break the time evolution up into steps of size $\Delta t$, with the $n$th step describing evolution from $t_{n-1}=(n-1)\Delta t$ to $t_n = n \Delta t$. During the $n$th time step, the Hamiltonian is independent of time and given by $\sum_A J_A(n) O_A$. The couplings $J_A(n)$ are Gaussian random variables with zero mean and covariance
\begin{equation}
    \Expval{J_A(n) J_{A'}(n')} = \frac{\delta_{nn'} \delta_{AA'}}{12(N-1)\Delta t} .
\end{equation}
The corresponding time evolution for the $n$th step is 
\begin{align}
    U(t_{n-1}, t_n) = \exp\left\{-i \Delta t \sum_A J_A(n) O_A \right\}.
\end{align}
The limit $\Delta t \to 0$ realizes the Brownian couplings in Eq.~\eqref{eq:bc_covar}.

In the case of the dressed OTOC, the governing Hamiltonians for the two branches (forward and backward) differ by a perturbation,
\begin{align}
    \dress{C}_{W,V} = \expval{\comm{\dress{W}(t)}{V}^\dagger \comm\Big{W(t)}{V}},
\end{align}
 where $\dress{W}(t)$ is the Heisenberg picture of an operator $W$ evolved by a perturbed Hamiltonian
\begin{align}
    \dress{H} = \sum \dress{J}_A(t) O_A,
\end{align}
where $\dress{J}_A \equiv (1-p)J_A + pX_A$. Here, $J_A$ and $X_A$ are sampled independently from an identical Gaussian distribution. $J_A$ and $\dress{J}_A$ have mutual covariance Eq.~\eqref{eq:forw_back_covar}, which relates the perturbation strength $p$ to the correlation parameter $r = (1-p)/\sqrt{1-2p+2p^2}$ used in subsequent equations. We assume that $W$ and $V$ are Pauli operators, in which case  
\begin{align}
    \dress{C}_{W,V} = \expval{2\dress{W}(t)W(t) - 2\dress{W}(t)VW(t)V}.\label{eq:dressed OTOC square commutator and 4-pt}
\end{align}
When $r=1$, this equation recovers the ideal OTOC case. To clarify the relationship between different OTOC definitions, we note that, in the ideal case, the squared commutator $C_{\text{commutator}}$ and the four-point correlator $C_{\text{correlator}}$ are related by
\begin{equation}
    C_{\text{commutator}} = 2 - 2C_{\text{correlator}},
\end{equation}
where $C_{\text{correlator}} = \langle W(t)VW(t)V \rangle$. In the dressed case, the first term on the RHS is the echo in Eq.~\eqref{eq:echo_kappa=0} and decays over time (to include noise, see Eq.~\eqref{eq:echo} with the full protocol where $\kappa\neq 0$). The dressed OTOC is $\dress{C}_{\text{commutator}}=\dress{C}_{W,V} = 2\times \text{echo} - 2\dress{C}_{\text{correlator}}$. The expression  explains why it is useful to divide the dressed OTOC by the echo, so that the resulting ROTOC can have a meaningful comparison with the physics of the ideal OTOC.

To compute the dressed OTOC and the echo, we expand $W,\dress{W}$ in the Pauli basis, $\dress{W}(t) = \sum_P \dress{c}_P(t)P$ and $W(t) = \sum_P c_P(t) P$, and take averages of the observables over random instances of $J_A$ and $J_{A'}$. Pauli strings with non-zero weight are traceless, so the echo becomes
\begin{equation}
     \Expval{\expval{\dress{W}(t)W(t)}}  = \sum_{P}  \Expval{c_P \dress{c}_P}.
\end{equation}
The dressed OTOC becomes 
\begin{equation}
    \dress{C}_{W,V} = 2 \sum_P c_P \dress{c}_P \big(1 - \expval{P V P V }\big).
\end{equation}
The factor $\langle P V P V \rangle$ is $1$ if $P$ and $V$ commute and $-1$ if they anti-commute. Taking random averages of $J_A$ and $J_{A'}$ gives 
\begin{align}
    \Expval{\dress{C}_{W,V}} = \sum_{P|\{P,V\}=0} 4 \Expval{c_P \dress{c}_P}.
\end{align}
Hence, after ensemble average, the echo and dressed OTOC are both related to the weight profile of the operator coefficients between the time evolution branches, which we denote as $\Expval{c_P \dress{c}_P}$.

Using the statistical symmetry of the Brownian cluster ensemble, it follows that the ensemble averaged weight profile does not depend on the precise makeup of the Pauli string $P$. It depends only on the total weight of the Pauli string (the number of non-identity Pauli's), so it is convenient to define the weight profile $b_w$ between branches at a given weight $w$:
\begin{align}
    b_w \equiv\sum_{P|\text{wt}(P)=w} \Expval{\dress{c}_Pc_P}.
\end{align}
Assuming $V$ has weight one, a short counting exercise shows that (details see Eq.~\eqref{eq:dotoc_bw_relation})
\begin{align}
    \Expval{\dress{C}_{W,V}(t)} = \sum_w \frac{8w}{3N}b_w(t),
\end{align}
and the ensemble averaged echo is
\begin{equation}
    \Expval{\expval{\dress{W}(t)W(t)}} = \sum_w b_w(t).
\end{equation}



There is one important exception to this statistical symmetry of the overlaps: the overlap with the initial operator which does not live in the permutation invariant space indexed by the weight $w$. This overlap is special because it has a non-zero average value which decays exponentially away from its initial value of unity (see \eqref{eq:autocorrelator_app} in App.~\ref{app_master_derivation}). For the conventional OTOC and the associated weight dynamics, this memory of the initial condition is rapidly lost and the assumption of statistical symmetry is a good one after a short time. For the renormalized OTOC, the inclusion of this special amplitude does potentially affect the form of the echo, $\Expval{\expval{\dress{W}(t)W(t)}}$, which receives two contributions, one from the special overlap with the initial operator and one from the statistically symmetric part of the weight profile. The special overlap with the initial operator decays at a rate $w_0$ and this overlap is squared in the echo, so it contributes a decay of rate $2w_0$. When $r=0$, this decay is precisely compensated by a growth in the rest of the weight profile, as required to have a constant echo. However, when $r<1$ the echo decays at a rate given by $2(1-r)$ times an averaged weight determined from the weight profile (see Sec.~\ref{subsec:methods_dyn}). This averaged weight is described in detail below, but its decay rate is  less than or equal to $w_0$ in the dilute limit. Hence, the decay into the statistically symmetric part of the weight profile is faster than the decay of the echo, and the assumption of statistical symmetry should remain valid, especially for $r$ close to unity.


\subsection{Operator dynamics}
\label{subsec:methods_dyn}

All that remains is to obtain the governing equation of the $b_w$'s. The Brownian cluster model provides two key advantages: first, both the dressed OTOC and echo can be represented in terms of $b_w$, and second, we obtain a closed dynamical equation for $b_w$ where the time evolution depends only on the $b_w$ values themselves, not on non-diagonal terms like $\Expval{c_P \tilde{c}_{P'}}$ with $P \neq P'$. The ensemble average over the Brownian cluster model enables the analytic expression of this dynamical equation. We defer the full derivation to App.~\ref{app_master_derivation}; including the effects of depolarization (at rate $\kappa$), the result is
\begin{widetext}
\begin{align}
    \dv{b_w}{t} =  -\frac{2w \Bigl((w-1)+3(N-w)\Bigr)}{3(N-1)} b_w - 2 w \kappa b_w  + 2r\qty[\frac{(N-w+1)(w-1)}{N-1}b_{w-1} + \frac{w(w+1)}{3(N-1)}b_{w+1}].\label{eq:b_w_eom}
\end{align}
\end{widetext}
When $r=1$ and $\kappa=0$, Eq.~\eqref{eq:b_w_eom} is a master equation in the sense that the $b_w$'s are all positive and $\sum_w b_w$ is conserved by the dynamics. In this full unitary case, the weight profile $\{b_w\}_w$ can be interpreted as a probability distribution. When $\kappa > 0$, terms such as $-2w\kappa b_w$ decrease the total weight profile $\sum_w b_w$, representing probability loss through decoherence. Similarly, when $r<1$, Eq.~\eqref{eq:b_w_eom} does not conserve $\sum_w b_w$, though it preserves the non-negativity of the $b_w$. This can be verified by noting that $\partial_t{b}_w\geq 0$ whenever $b_w$ reaches $0$ from a positive value, preventing any component from becoming negative.

More explicitly, we can rewrite Eq.~\eqref{eq:b_w_eom} in terms of a rate matrix $M$, 
\begin{align}
    \dv{b_w}{t} &= (M b)_w. \label{eq:rate_matrix_bw}
\end{align}
$M$ has the form
\begin{align}
    &M_{w,w'}(r,\kappa) = r M_{w,w'}(1,0) - 2w \kappa \delta_{w,w'} \nonumber \\
      & - (1-r) \frac{2w \Bigl((w-1)+3(N-w)\Bigr)}{3(N-1)} \delta_{w,w'}.
\end{align}
$M(1,0)$ is a probability-conserving stochastic rate matrix corresponding to the ideal OTOC evolution. Likewise, $M(r,\kappa)$ can be interpreted as rescaling the evolution rate by $r$ and adding two probability loss terms proportional to $1-r$ and $\kappa$. We can also see that the evolution preserves positivity, since the $M(1,0)$ preserves positivity and the extra loss terms vanish when $b_w$ approaches zero.

Since the echo is the total weight profile $\sum_w b_w$, it captures the loss of weights when $r<1$ or $\kappa>0$. Dividing the dressed OTOC by the echo amounts to renormalizing the $b_w$ distribution, so the ratio of the dressed OTOC to the echo is equal to
\begin{equation}
    \frac{\Expval{\dress{C}_{W,V}}}{\Expval{\expval{\dress{W}(t)W(t)}}} = \sum_w \frac{8w}{3N}c_w(t) = \frac{8 \langle w\rangle_c}{3N} \label{eq:rotoc_avg_weight}
\end{equation}
where
\begin{equation}
\label{eq:c_w_t}
    c_w(t) = \frac{b_w(t)}{\sum_{w'} b_{w'}(t)}
\end{equation}
is a normalized probability distribution and $\langle \cdots \rangle_c$ denotes the average weight under this distribution. This result illustrates that the ROTOC is a particularly natural observable in the Brownian cluster model, since it directly measures the average weight of a proper probability distribution that remains well-defined even in the presence of perturbation and decoherence.

We can also obtain an evolution equation for the conserved distribution $c_w$. Taking the time derivative of Eq.~\eqref{eq:c_w_t} and using Eq.~\eqref{eq:rate_matrix_bw}, we obtain in vector matrix notation: 
\begin{equation}
     \dv{c}{t} = M c + \mu c \label{eq:c_eom}
\end{equation}

where
\begin{equation}
    \mu = - u^T M c
\end{equation}
and $u$ is a vector of all $1$s. Since $M$ decreases the total probability mass when $r<1$ or $\kappa>0$, it follows that $\mu \geq 0$. The $\mu c$ term in Eq.~\eqref{eq:c_eom} has the interpretation of adding back the lost probability. It is also worth noting that Eq.~\eqref{eq:c_eom} is non-linear in the $c$ variables.

\section{Results\label{sec:results}}



\begin{figure}[t]
    \centering
    \includegraphics[width=\linewidth]{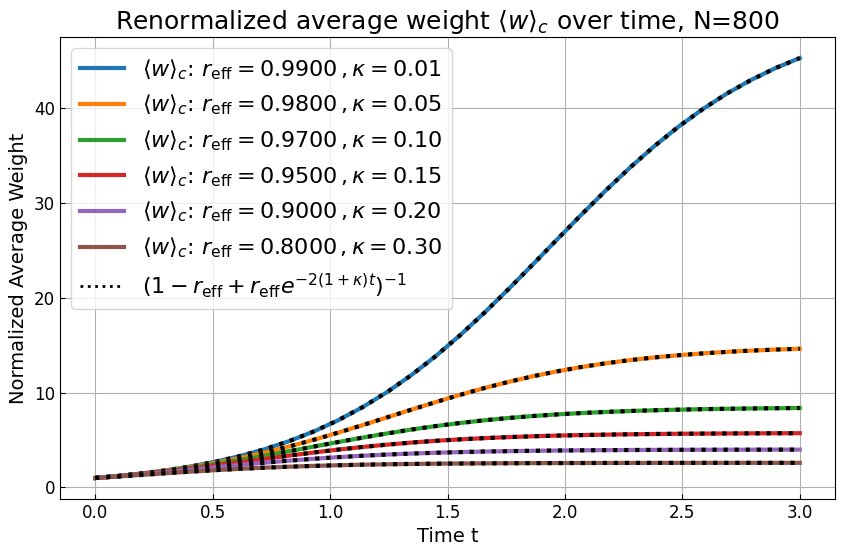}
    \caption{Average weights $\expval{w}_c$ under the renormalized probability distribution $c_w$ with a selection  of perturbation strengths ($r$, or the rescaled $r_{\text{eff}}=r/(1-\k)$) and noise rates ($\k$) in a system of $N=800$ qubits are plotted over time for the dilute limit dynamics Eq.~\eqref{eq:b_w_dilute_M}. Here, we compare the dilute limit numerics (solid lines) with the analytical prediction (dotted lines). }
    \label{fig:renormalized}
\end{figure}

In this section, we present solutions of Eq.~\eqref{eq:b_w_eom} and discuss the physics of the ROTOC in various limits. The Brownian cluster model reduces the ROTOC problem to solving an ordinary differential equation in discrete weight space, which has polynomial computational complexity in terms of the number of spins. Since we have reduced the complexity of the quantum dynamics to $N$ numbers, $b_1, \cdots, b_N$, we can access much larger systems than would be possible with exact quantum calculations. An example calculation with $N=800$ qubits is shown in Fig.~\ref{fig:renormalized}, where the dashed lines represent the exact formulas derived in the dilute limit in Sec.~\ref{subsec:results_dilute}.

The $N \rightarrow \infty$ limit greatly simplifies the evolution equations of $b_w$ and $c_w$, and we present the exact solutions at the leading order below. At order $1/N$, we analyze the lifetime of metastable states that emerge in the dynamics (see Figure~\ref{fig:metastable}). The qualitative shape of the curves in Fig.~\ref{fig:renormalized} is consistent with experimental data from \cite{Alvarez2015}, and we discuss this comparison further in Sec.~\ref{subsec:results_nmr}.

\subsection{Dilute limit}
\label{subsec:results_dilute}

We first consider the dilute limit in which the typical size of the Pauli strings (measured by $\langle w\rangle_c$) is much smaller than $N$. Mathematically, it corresponds to taking $N \rightarrow \infty$ while keeping $r \in [0,1)$ and $\k \ge 0$ fixed. At the leading order, the equation of motion Eq.~\eqref{eq:b_w_eom} of $b_w$ is simplified to
\begin{equation}
    \dv{b_w}{t} =  - 2 w(1+\kappa) b_w  + 2 r (w-1) b_{w-1} .\label{eq:b_w_dilute_M}
\end{equation}

Note that we can remove $\kappa$ from the equation by a redefinition of  $r$ and a rescaling of  time $t$ to effective variables: 
\begin{eqnarray}
    r_{\rm eff} &= \frac{r}{1 + \kappa}, \label{eq:r_eff}\\
    t_{\rm eff} &= (1 + \kappa) t.
\end{eqnarray}

Specifically, we can express the dilute limit rate matrix in the form
\begin{equation}
    M(r,\k)_{w,w'} = - 2 w(1+\k) \delta_{w,w'} + 2 r (w-1) \delta_{w-1,w'}.\label{eq:dilute_M}
\end{equation}
By extracting a factor of $(1+\k)$, we can write
\begin{equation}
    M(r,\kappa)=(1+\kappa)M(r_{\text{eff}},0),
\end{equation}

Hence, up to a rescaling of time, it is sufficient to study the $r$ dependence with $\kappa=0$. Consistent with this, $\mu$ with the dilute rate matrix Eq.~\eqref{eq:dilute_M} is given by
\begin{equation}
    \mu = 2(1+\kappa-r) \langle w\rangle_c = (1+\kappa)2(1-r_{\text{eff}})\langle w\rangle_c,
\end{equation}
where again $\langle g(w) \rangle_c = \sum_w c_w g(w)$ is the normalized expectation of function $g(w)$ according to the $c_w$ weight distribution.

Therefore to simplify the discussion, we will set $\kappa = 0$ in the following. Eq.~\eqref{eq:b_w_dilute_M} becomes
\begin{equation}
    \dv{b_w}{t} = -2wb_w + 2r(w-1)b_{w-1}\label{eq:bw_kappa=0}.
\end{equation}
There is only a process of increasing the weight from $w-1$ to $w$, the weight decreasing process is killed at this order. 

However, due to the presence of $r$, there is a rate of $2w - 2rw = 2w( 1  -r )$ to decrease the total weight. The $w$ in the coefficient indicates that larger weight decreases faster. This is evident from the equation for the probability distribution $c_w$:
\begin{equation}
    \dv{c_w}{t} =  - 2 w c_w  + 2 r (w-1) c_{w-1}  + 2(1-r) \langle w\rangle_c c_w 
    \label{eq:c_w_dilute_M}
\end{equation}
where again $\langle \cdots \rangle_c$ is an average over the distribution $c_w$. This form shows clearly how the final non-linear term adds back the lost part of the weight profile, making $c_w$ a properly normalized probability distribution at all times. 

In App.~\ref{app:master_dilute}, we derive the self-consistent equation for the generating function  $ C(z, t) = \sum_{w= 0}^{\infty} c_w z^w$  for the initial condition where the operator concentrates on one specific weight $w_0$, which translates to $b_w = c_w = \delta_{w , w_0}$ 
 \begin{align}
     C(z, t) &=  \exp\left\{ \int_0^t ( 1 - r) \langle w(\tau) \rangle_c d \tau \right\} \nonumber\\
     &\hspace{2cm} \times \left(  \frac{z}{1 - r z( 1 - e^{-t } ) }\right)^{w_0} .
 \end{align}
The solution of the average weight can be found easily from the generating function 
\begin{equation}
    \langle w(t) \rangle_c = \frac{w_0}{1 - r + r e^{ - t} }.
    \label{eq:w_t_dilute_sol}
\end{equation}

When $r \gg 1 - r$, (more precisely $r_{\text{eff}} \gg 1 - r_{\text{eff}}$ when $\kappa >0$ ) we can see an early time exponential growth
\begin{equation}
    \langle w\rangle_c = \frac{w_0 e^{2(1+\kappa)t}}{1-r_{\text{eff}}} + \dots \,\,\,\, ( \text{early time}).
\end{equation}
At late times, the average weight saturates to 
\begin{equation}
    \langle w\rangle_c \to \frac{w_0}{1-r_{\text{eff}}} \,\,\,\, (\text{late time}).\label{eq:late_time_w_c}
\end{equation}
However, these solutions ($w_0 >1$) are metastable states with an infinite lifetime only when $N$ is strictly infinite. When $N$ is large but finite, there will be weight decreasing processes at order $1/N$, and the state with $\langle w\rangle_c = \frac{w_0}{1-r_{\text{eff}}}$ is eventually unstable, as we discuss in the next subsection.

For the specific case of initial condition $c_w=\delta_{w,1}$ (i.e., $w_0=1$), the weight dynamics simplifies to
\begin{equation}
    \langle w\rangle_c(t)=\frac{1}{1-r_{\text{eff}}+r_{\text{eff}} e^{-2(1+\kappa)t}}.
\end{equation}

Including the $1/N$ normalization from Eq.~\eqref{eq:rotoc_avg_weight}, the ROTOC in the Brownian cluster model is
\begin{equation}
    \text{ROTOC} = \frac{8}{3N} \frac{1}{1-r_{\text{eff}}+r_{\text{eff}} e^{-2(1+\kappa)t}}.
\end{equation}
When $\kappa=0$, we thus find a remarkable result that the ROTOC exhibits exponential growth at exactly the unperturbed rate, although it saturates to a value set by $r<1$ rather than by $N$. Even when $\kappa>0$, the growth rate of the ROTOC is renormalized by a definite factor, so the $\kappa=0$ rate can still be extracted from the $\kappa>0$ data.

\subsection{Metastable states}
\label{subsec:results_metastable}


The result in Eq.~\eqref{eq:w_t_dilute_sol} shows that there is an infinite family of stable states indexed by the initial operator weight when $N \rightarrow \infty$. However, this behavior only exists in the strict dilute limit where it arises because all the terms in the equations of motion that produce a flow of operators to lower weight are $1/N$ suppressed. Numerical integration of the full equations of motion at large but finite $N$ shows that these states are only metastable. This can be seen directly from \eqref{eq:b_w_eom}: the only term which moves the weight profile $b_w$ from higher weight to lower weight is,
\begin{equation}
    r \frac{2 w(w+1)}{2(N-1)} b_{w+1},
\end{equation}
which is $1/N$ suppressed in the dilute limit with $w \ll N$. Hence, when this term is dropped, it is impossible for the system to move to lower weights. 

Numerical results indicate a finite lifetime of these metastable states that scales as $\ln N$ when $w_0 \ll N$. 
We can understand the finite lifetime of the metastable state by re-including the effects of the suppressed operator weight decreasing term. The key idea is to estimate the ratios of the probabilities $c_{w_0}$ and $c_1$. At the metastable state, $ c_{1} = 0$, but once $c_1$ has a small non-zero value, its magnitude grows exponentially relative to $c_{w_0}$ because the reweighting term in \eqref{eq:c_w_dilute_M} far outstrips the natural decay of $c_1$. 
As the two terms $c_{w_0}$ and $c_1$ evolve, they eventually become comparable, $c_1 \sim c_{w_0}$, and it is this time we take as an estimate of the lifetime. 

We now perform the estimation in the regime  $w_0 \ll N$ and $1 - r \ll 1$ for simplicity. 
Suppose the system has reached the metastable state ignoring the $1/N$ suppressed terms, so that $w_0$ is the lowest index of which $c_w$ is non-vanishing. Now turn on the $1/N$ operator weight down flow. 
From  Eq.~ \eqref{eq:b_w_eom}, we see that $c_{w_0-1}$ will be sourced from $c_{w_0}$. Balancing the inherent decay of $c_{w_0-1}$ with the down flow from $c_{w_0}$, we get an estimate
\begin{equation}
    -2 (w_0-1) c_{w_0-1} + \frac{2 r w_0 (w_0-1)}{3N}c_{w_0} \approx 0,
\end{equation}
which gives
\begin{equation}
    c_{w_0-1} \approx \frac{r w_0}{3N} c_{w_0}.
\end{equation}
Iterating this formula, we estimate that
\begin{equation}
    c_1 \approx \frac{(w_0)!}{(3N/r)^{w_0}}.
\end{equation}
We thus have an estimate for $c_1$ established by the early time dynamics. 

Next we suppose the reweighting process takes over, which corresponds to the last term on the right of Eq.~\eqref{eq:c_w_dilute_M}.  The metastable state is losing probability mass rapidly at a rate $-2 (1-r) \langle w\rangle \sim - 2 w_0$, whereas the true stable state only loses probability mass at a rate $-2$. Hence, there is a rapid transfer of probability mass from the metastable state to the true stable state. However, this exponential growth of the true stable state must overcome the small initial condition. This leads to a lifetime $\tau$ obeying
\begin{equation}
    e^{2 (w_0 - 1) \tau} \frac{(w_0)!}{(3N/r)^{w_0}} \sim 1.
\end{equation}
Solving for $\tau$ gives
\begin{equation}
    \tau \sim \frac{w_0}{2(w_0-1)} \ln \frac{3 e N}{r w_0}. \label{eq:metastable_lifetime}
\end{equation}
For example, with $1-r \ll 1$ and $w_0 \gg 1$, the analysis predicts a lifetime $\tau \sim \frac{1}{2}\ln \frac{N}{w_0}$.

We note that there can be other metastable states. For example, the state of a completely random operator can occur if $w_0 \sim N$ and $1 - r \ll 1$. A similar reweighting mechanism  shifts the metastable state to a true stable state, and the analysis above can in principle be carried out. However, it is harder to make a precise prediction and the limit of $w_0 \sim N$ is also less relevant in the experiment. 

To probe the $N$ dependence in numerics, we use a trick to decouple the parameter $N$ as the number of weights --- thus relating to the numerical complexity --- and $N$ controlling the $1/N$ coefficients. In the dilute limit,  the maximum weight reached by the dynamics is controlled by $w_0$ and $r$, and the magnitude of $c_{w_0+\ell} \sim r^\ell$ decays exponentially in the metastable state. We can therefore truncate the number of weights to be $w \le N_{\rm cut}$  and set the parameters controlling the coefficients of the equation to be $1/N_{\rm eff} $. 

\begin{figure}
    \centering
    \includegraphics[width=\linewidth]{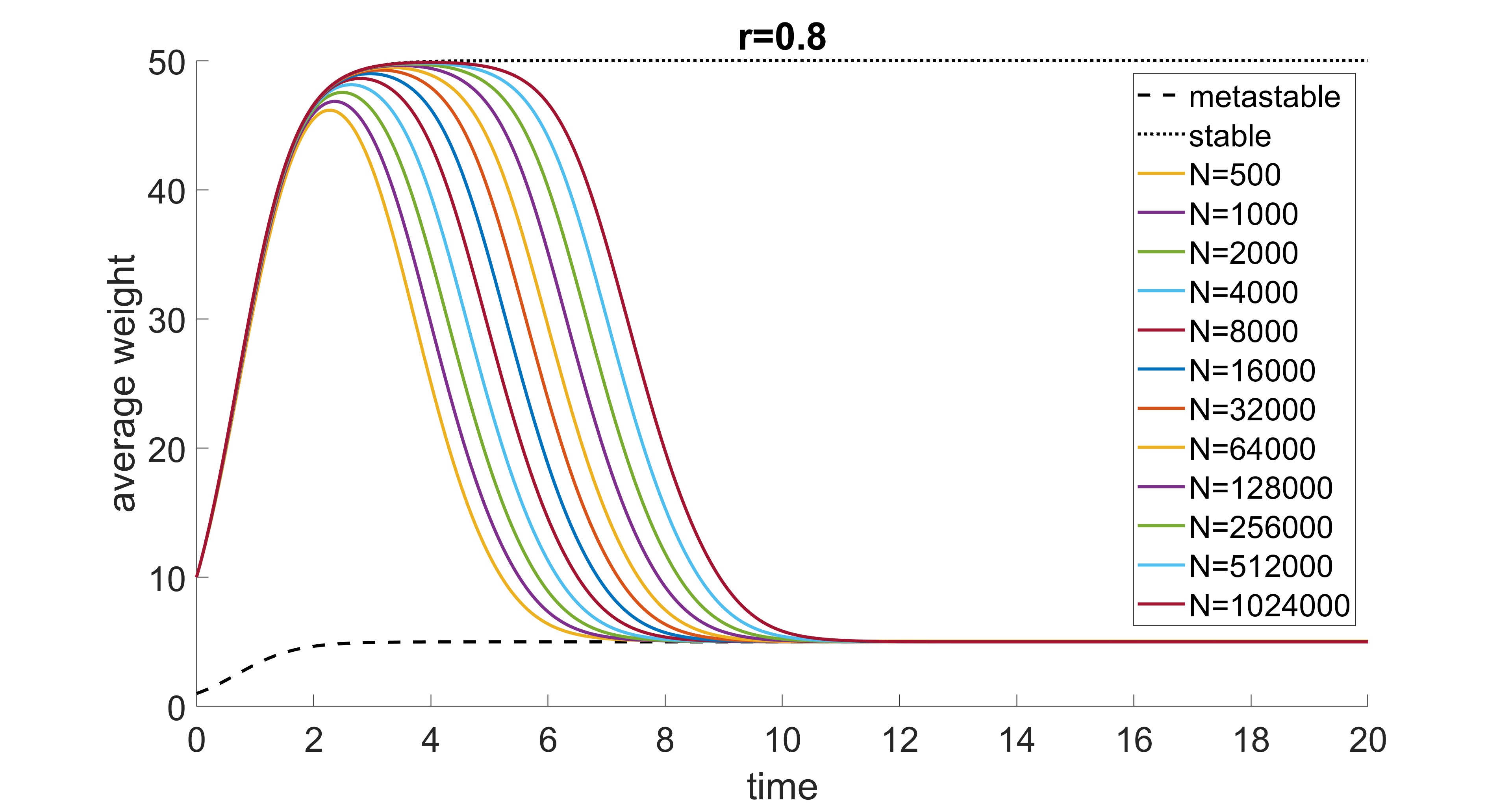}
    \caption{Average weight $\langle w \rangle_c$ versus time from a numerical solution of the equations of motion Eq.~\eqref{eq:c_eom}. We set $N_{\text{cut}}=500$ to be the maximal weight, $r=.8$, $w_0=10$, and $N_{\text{eff}}=N_{\text{cut}} 2^\ell$ for $\ell=0,\cdots, 11$ to  be the variable $N$ in the coefficients of the equations of motion.
    The choices of $N_{\text{eff}}$ increase by a constant factor from curve to curve, so the $\log N_{\text{eff}}$ predicted lifetime translates into a constant shift of the lifetime from curve to curve. This matches well with the numerical data. The order of magnitude of the lifetime is also accurately captured by \eqref{eq:metastable_lifetime}. }
    \label{fig:metastable}
\end{figure}

We can then effectively probe large $N_{\text{eff}}$ while keeping $N_{\text{cut}}$ fixed provided we work in the dilute limit. Numerical data supporting the above estimate are shown in Fig.~\ref{fig:metastable}. We have also verified the logarithmic dependence of $\tau$ on $w_0$ (not shown).

\subsection{Comparison with nuclear spin dynamics}
\label{subsec:results_nmr}

One important motivation of our work is to explain OTOC measurements with noise and imperfections in the nuclear spin data. In this subsection, we compare our results of ROTOC with the measured data of  \cite{Alvarez2015} and the ``localization-like" phase transition in the parameter space of perturbation and decoherence. 
Before going into the analysis, we first discuss the key differences between the experimental setup in \cite{Alvarez2015} and our setup. 

First, note that the interactions of our model and the Hamiltonian of the adamantane material used in \cite{Alvarez2015} are not identical. For theoretical simplicity, we use all-to-all random interactions, where the interaction strength does not decay in distance and the couplings are time-dependent, while \cite{Alvarez2015} studies a system modeled by a time-independent Hamiltonian with dipolar interactions. Our model captures the chaotic and long-range features of the experiment and, in a sense, pushes them to the limit. Generally speaking, such a limit enhances the scrambling despite the introduction of decoherence. Second, the OTOC considered in \cite{Alvarez2015} is built from global spin operators. In Ref.~\cite{Zhou2023a} it was argued that the OTOC of global spin operators reduces to a sum over local OTOCs for sufficiently chaotic systems, a claim supported with finite system size numerical evidence in a 1D spin chain with power-law interactions; this idea was substantiated in the context of nuclear spins with a distinct set of results in Ref.~\cite{Lozano_Negro_2024}. A sum of local OTOCs corresponds to $\expval{w_c}$, which is what we compare to the ``cluster size'' data in \cite{Alvarez2015}.

To summarize, first, although the models are different, they plausibly have rapid and roughly comparable operator growth in the unperturbed limit. Second, the right quantity to compare is the average weight $\expval{w}_c$ without any $1/N$ normalization. 

Our model can reproduce the rapid early time growth in the extended phase of  Ref.~\cite{Alvarez2015}, although the detailed functional form of the average weight as a function of time is different. In fact, we showed that the Lyapunov exponent is $(1 + \kappa )\lambda$ when $\kappa \ne 0$ and $(1-r) \ll r$. However, we don't see a phase transition --- our model does not host a delocalized phase at a sufficiently long time. With the possibility of going through a meta-stable state, the system will eventually settle down to a steady state where the average operator weight is $O(N^0)$. Since our model enhances scrambling by taking random all-to-all interactions, the absence of a delocalization phase transition suggests that no such phase transition should exist in the real adamantane model. We instead propose that the observations in \cite{Alvarez2015} correspond to a rapid crossover as a function of $r$.

In terms of parameters, the setup in \cite{Alvarez2015} is presumably studying a large system ($N \to \infty$) at a relatively short time. We model the experiment as having a fixed value of $\kappa$ and variable $r$ corresponding to their adjustable ``perturbation strength''; in App.~\ref{app:dressed_otoc}, we discuss the mapping between these parameters. Since $\kappa$ is fixed, we can absorb it into the unit of time and study the dependence on $r_{\text{eff}}$. This is equivalent to simply setting $\kappa=0$, which we do in the remainder of this subsection.

Since $N$ is taken to be large, we can work in the dilute limit. The results of Sec.~\ref{subsec:results_dilute} then immediately show that the global ROTOC shows no true phase transition in the thermodynamic limit in our model. This is in contrast to the evidence for a phase transition as a function of perturbation strength presented in \cite{Alvarez2015}. As noted above, given the differences in our models, we cannot rule out that their setup has a genuine phase transition even though ours does not. However, we think this is unlikely and would instead propose that the observations in \cite{Alvarez2015} correspond to a rapid crossover as a function of $r$.


{\begin{figure}[t]
    \centering
    \includegraphics[width=1\linewidth]{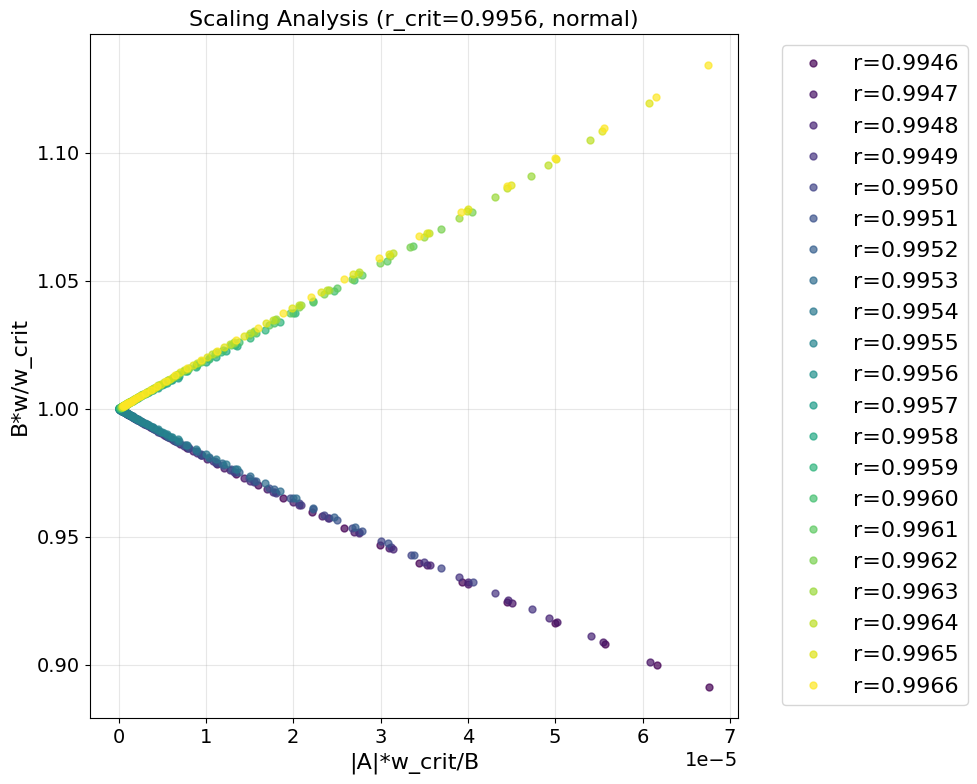}
    \caption{Scaling collapse of operator growth dynamics revealing two distinct branches. The critical parameter $r_{\text{crit}} = 0.9956$ presented here is determined according to the definition introduced in Appendix~\ref{app:dilute_ftc}. However, it is important to note that this collapse is not dependent on the specific value of $r_{\text{crit}}$; additional numerical results for other $r_{\text{crit}}$ values are presented in the App.~\ref{app:numerics}, where we observe the same kind of collapse into two branches. Here, the cutoff time $T$ is determined by the time point at which $\partial_t^2 \langle w \rangle_c (r=1, t)$ reaches maximum value. This ensures that our analysis remains within the dilute limit regime appropriate for drawing parallels with the NMR experiment \cite{Alvarez2015}. We employ $w_{\text{crit}} = \langle w \rangle_c (r=r_{\text{crit}}, t)$ as our time variable, with scaling parameters $A=1-r/r_{\text{crit}}$ and $B=r/r_{\text{crit}}$. }
    \label{fig:scaling_0.9956}
\end{figure}

To model such a crossover, we fix a maximal evolution time $T$ in the experiment and an associated maximum average weight $w_{\max} = \langle w\rangle_c^{r=1}(t=T) = e^{2 T}$.  We define the crossover $r_*$ via
\begin{equation}
    \partial_t^2 \langle w \rangle^{r=r_*}_c(T) = 0, \label{eq:inflection_point}
\end{equation}
i.e. the $r$ for which there is an inflection point at time $T$. For $r>r_*$, the average weight is still accelerating at time $T$, while for $r<r_*$, the average weight has begun decelerating by time $T$. When $T$ is large, so that the maximum average weight is large, the crossover point is 
\begin{equation}
    r_* \approx 1 - \frac{2}{w_{\max}}
\end{equation}
as shown in App.~\ref{app:dilute_ftc}.

The $r$ dependence of $\langle w \rangle_c(t)$ then roughly reproduces the perturbation strength dependence of the cluster size data in \cite{Alvarez2015}: for $r>r_*$, one sees unbounded growth over the given time window, while for $r<r_*$, the growth has begun to slow and level off. The detailed scaling analysis discussed in \cite{Alvarez2015} does not appear compatible with our model, however, the order of magnitude of the perturbation strength required to see a crossover is consistent with our theory. Eq.~\eqref{eq:r_p_relation} relates our $r$ quantity to an analog perturbation strength, $p$, considered in \cite{Alvarez2015}. When $r$ is close to one, we find $1-r = \frac{p^2}{2} + \cdots$. Taking $w_{\max} \sim 5 \times 10^3$ as an estimate of the maximum average weight from the $p=0$ data in \cite{Alvarez2015}, we find $1-r_* = 4 \times 10^{-4}$ and $p_* \approx .028$ which is roughly comparable to the $p_c \approx .027$ reported in \cite{Alvarez2015}. The precise value will depend on the details, for example, the $p=0$ data in \cite{Alvarez2015} still presumably has $\kappa>0$ and hence $r_{\text{eff}}<1$ and there is a factor of $8/3$ in going from the average weight to the global ROTOC, but the order of magnitude appears reasonable. The fact that the effect of the perturbation grows with $p^2$ rather than with $p$ is known in other contexts as well, e.g. \cite{Jalabert_2001}, and it plausibly holds even for systems with a fixed (non-random) Hamiltonian provided it is sufficiently chaotic so that the perturbation behaves ``incoherently''.  

In addition, we give a scaling collapse analysis for our system. For simplicity, consider $w_0=1$ and a critical correlation strength $r_{\text{crit}}$. Rewrite the dilute limit solution 
\begin{equation}
    e^{-2t}=\frac{1}{wr}-\frac{(1-r)}{r},
\end{equation}
we can then use $w(r=r_{\text{crit}}, t)$ as the time variable, and rewrite the average weight solution in dilute limit,
\begin{equation}
    w(w_\text{crit}, r)=\frac{1}{A+\frac{B}{w_\text{crit}}}.
\end{equation}
where $A\equiv1-\frac{r}{r_\text{crit}}$ and $B\equiv\frac{r}{r_\text{crit}}$. We further rearranged this to 
\begin{equation}
    \frac{Bw}{w_\text{crit}}=\frac{1}{1+\frac{Aw_\text{crit}}{B}}.
\end{equation}
Plotting the LHS with respect to $\frac{\abs{A}w_\text{crit}}{B}$ gives the collapse as illustrated in Fig.~\ref{fig:scaling_0.9956}, where we choose $r_{crit}=r_*$ for an $N=2000$ system. However, we note importantly that the collapse depends weakly on the choice of $r_\text{crit}$, and the above analysis would result in collapse into two branches regardless of $r_\text{crit}$ picked. Two more examples with scaling collapse are plotted in the App.~\ref{app:numerics} for arbitrarily picked $r_{\text{crit}}=0.9862, 0.9988$ as evidence for the absence of a phase transition in our system.

\section{Outlook\label{sec:outlook}}

Motivated by numerous issues raised by existing and future scrambling experiments \cite{Alvarez2015, Dominguez2021, Dominguez2021a}, we have studied imperfections in scrambling dynamics in a solvable model. After averaging over random couplings, we formulated the full operator growth dynamics as a set of non-linear equations of motion, thus reducing the complexity from tracking $\sim 4^N$ operator amplitudes to tracking $\sim N$ probabilities of weights. Furthermore, in the dilute limit obtained by taking the thermodynamic limit with fixed non-vanishing imperfections, we were able to obtain a simple analytic expression for the average weight dynamics. With this result, we were able to conclusively show that there is no ``localization'' phase transition in the thermodynamic limit, although we did find evidence of a rapid crossover with characteristics potentially compatible with data from ensembles of nuclear spins.

A key result of our study is that the renormalized OTOC studied in \cite{Swingle2018} indeed sticks close to the unperturbed dynamics at early time provided the perturbation is not too strong. Moreover, we have given explicit forms of the renormalized OTOC which allow us to extract the unperturbed chaos exponent given a knowledge of dissipation strength $\kappa$. Even knowing the precise value of $r$ is not required, provided both $1-r$ and $\kappa$ are small enough that a significant period of exponential growth can be observed. The detailed control we have over the model and the subtle differences between the effects of perturbations and decoherence also suggest that this model can be very useful as a benchmarking tool in scrambling studies. 

Several important questions remain for future work. One key direction concerns the generality of our results. Do they extend qualitatively or quantitatively beyond the Brownian cluster model? For example, we believe that the absence of a phase transition in the thermodynamic limit is generic, and further evidence in a more realistic model would strengthen this claim. Investigating whether weak perturbations can create binding transitions similar to those in \cite{Jacoby2024} would be valuable. Examining the effects of other decoherence types is also worthwhile, though many channels may produce similar effects when different Pauli strings have statistically similar amplitudes, as commonly occurs in chaotic models. The influence of model-specific features, such as symmetries and geometric locality \cite{Khemani2018,Cheng2021}, merits further exploration.

A second direction involves extending this work to OTOCs at finite temperature \cite{Shenker2014, Green2022, Sundar2022, Schuster2022a}. While decoherence and perturbations may cause heating, relevant observables can still be formulated by fixing an initial state at a prescribed energy density, such as a Gibbs state, and implementing a generalization of the protocol in Fig.~\ref{fig:main_results}.

A third direction examines the implications of our results for quantum information protocols. For instance, how does the ROTOC relate to many-body teleportation schemes based on scrambling \cite{Yoshida2019, Brown_2023, Seki2024, Hayden2007, Nezami2023}? The consequences for metrological and state learning protocols involving OTOCs \cite{Li2023a, Schuster2023a, Davis2016, Holmes2021, Leone2024a, Kobrin2024, Garcia2021a, McGinley2022a} as well as OTOCs' roles in resource theory and circuit complexity \cite{Bu2024, Garcia2023b} also warrant investigation.

Finally, conducting a detailed study of this model or related models \cite{Yin2020} using quantum simulation platforms would be valuable---both to benchmark against our exact results and to explore their applicability to less analytically tractable models. It would be particularly exciting to investigate the metastability phenomenon we described. 

\textit{Acknowledgements:} We thank Douglas Stanford, Xiao-liang Qi, Ruoyi Yin, Valérie Bettaque, and Shiyong Guo for discussions. NL and BS gratefully acknowledge support for this work from the Heising-Simons Foundation under grant 2024-4849.




\bibliography{lbib.bib}
\bibliographystyle{quantum.bst}


\newpage
\appendix
\begin{widetext}

\section{Protocol including decoherence}
\label{app:protocol}

Here we describe the full protocol including explicit decoherence which generalizes the discussion in Sec.~\ref{subsec:methods_obs}. We model decoherence as a depolarizing channel that acts during the time evolution steps (the $U$ and $\dress{U}$ steps). For a single qubit, a depolarizing channel of strength $p$ acts as
\begin{equation}
    \rho \to D_p(\rho) = (1-p) \rho + p \frac{I}{2}.
\end{equation}
Expanding $\rho$ in the Pauli basis, $\rho = \frac{I + \vec{a} \cdot \vec{\sigma}}{2}$, $D_p$ maps 
\begin{equation}
    \vec{a} \to (1-p) \vec{a}.
\end{equation}
It is convenient to introduce a notion of operator weight, with the identity being weight zero and each Pauli operator being weight one. Then we can describe the depolarization channel succinctly as damping all the weight one operators by $(1-p)$ and leaving the weight zero operator alone.

For many qubits, independent depolarizing noise on each qubit has the effect of damping Pauli strings based on their weight. Expanding a many-body $\rho$ in terms of a complete set of Pauli strings $\{P_S\}$ as 
\begin{equation}
    \rho = \sum_S a_S P_S,
\end{equation}
we have
\begin{equation}
    D_p^{\otimes N}(\rho) = \sum_S a_S P_S (1-p)^{w(S)}
\end{equation}
where $w(S)$ is the weight of string $S$, which is the number of non-identity Pauli operators in the string.

We now discuss the full protocol with noise. It is convenient to describe this protocol using an infinitesimal version of the depolarizing channel; setting $p=\kappa dt$, we get
\begin{equation}
    \sum_S a_S P_S \to \sum_S a_S P_S e^{- \kappa w(S) dt}. \label{eq:op_decoherence}
\end{equation}
We also introduce a time $T$ and Hamiltonians $H$ and $\dress{H}$ which generate $U = e^{- i H T}$ and $\dress{U}= e^{- i \dress{H} T}$. The noisy protocol is:
\begin{enumerate}
    \item Prepare the initial state, $\rho = \frac{I + \epsilon V}{2^N}$;
    \item Evolve forward, $\rho \to D_{\kappa dt}^{\otimes N}(e^{- i H dt} \rho e^{i H dt}) $, repeated $T/dt$ times;
    \item Rotate, $\rho \to e^{i \phi W}\rho e^{-i \phi W}$;
    \item Evolve backward, $\rho \to D_{\kappa dt}^{\otimes N}(e^{ i \dress{H} dt} \rho e^{-i \dress{H} dt}) $, repeated $T/dt$ times;
    \item Measure $V$.
\end{enumerate}
Note that the dissipation acts the same way in both evolution steps, since it models physical processes beyond the experimenters' control.

Let $D_U$ denote the complete forward evolution and $D_{\dress{U}^\dagger}$ denote the complete backward evolution. The expected value of the final $V$ measurement now yields a generalization of $f_{\text{ideal}}$,
\begin{equation}
    f = \expval{ V D_{\dress{U}^\dagger}( e^{i \phi W} D_U (V) e^{- i \phi W})}.
\end{equation}
Since the adjoint of the depolarizing channel is itself, we arrive at 
\begin{equation}
    f = \expval{ D_{\dress{U}}(V) e^{i \phi W} D_U(V) e^{-i \phi W}}.
\end{equation}
The noisy version of the dressed OTOC is
\begin{equation}
    \partial_\phi^2 f(\phi)\big|_{\phi=0} = \langle [W,D_{\dress{U}}(V)][W,D_U(V)]\rangle,
    \label{eq:dressed_otoc_kappa}
\end{equation}
and the noisy echo is
\begin{equation}
 f(\phi)\big|_{\phi=0} = \expval{ D_{\dress{U}}(V) D_U(V) }.
 \label{eq:echo}
\end{equation}
The renormalized OTOC is then the ratio of these two quantities.

\section{Discrete Brownian Cluster Model\label{app_master_derivation}}
In App.~\ref{app:ordinary_otoc}, we discuss the motivation for using the discrete Brownian cluster model, and review the calculation of the ideal OTOCs in this model. In App.~\ref{app:dressed_otoc}, we generalize to the echo and the dressed OTOC in the presence of a perturbation represented by the correlation coefficient $r$ between the ``forward" ($J_A$) and backward ($\dress{J}_A$) branch of the time evolution described in Eq.~\eqref{eq:forw_back_covar}. The relation is reproduced here:
\begin{equation}
    \mathbb{E}[J_A(t) \dress{J}_{A'}(t')] = r \mathbb{E}[J_A(t) J_{A'}(t')]. 
\end{equation}
For both parts, we derive the master equations governing the operator dynamics in the Brownian model that will give rise to the dynamics of OTOC its generalization.
\subsection{Ordinary OTOC in the discrete Brownian cluster model\label{app:ordinary_otoc}}

The motivation for choosing the discrete Brownian cluster model is summarized below. The idea is to choose a model as simple as possible while still allowing us to probe beyond the symmetric subspace for a meaningful many-body phenomenon. For this, we would like a Hamiltonian that:
\begin{enumerate}
    \item Is as symmetric as possible among qubits: no qubit is special;
    \item Is as symmetric as possible among directions: no direction is singled out;
    \item Has statistical permutation symmetry among spins at different sites: no interaction is special;
    \item Has all conservation laws removed, so all that is left is information;
    \item Is randomized in dynamics to obtain a useful set of ensemble-averaged observables.
\end{enumerate}

One Hamiltonian that meets these criteria is the discrete Brownian cluster model
\begin{align}
    H = \sum_{i,j,\alpha, \beta} J_{ij}^{\alpha\beta}\sigma^\alpha_i\sigma^\beta_j
\end{align}
where the $i,j$ label denotes sites and $\alpha, \beta$ label denotes the non-identity Pauli's. We sample $J_{ij}^{\a\b}$ from Gaussian i.i.d so its value is independent of its indices. This satisfies the statistical permutation symmetry.

Finally, to satisfy the last two criteria, the Hamiltonian is made time-dependent, updating in every time step $\Delta t$ with a set of new coupling constants $J_{ij}^{\alpha\beta}$. The time evolution operator in a single time step is
\begin{align}
    U(t, t+\Delta t) = \exp{-i \Delta t \sum_A J_A O_A }
\end{align}
where we use $A$ to denote the multi-index $ij\alpha\beta$ and $O_A\equiv \sigma_i^\alpha\sigma_j^\beta$. The full time evolution operator is
\begin{align}
    U(0, t) &= \prod_{k=0}^{t/\Delta t - 1} U(k\Delta t, (k+1)\Delta t) = \prod_{k=0}^{t/\Delta t - 1}\exp{-i \Delta t \sum_A J_A O_A}.
\end{align}

Before we calculate the OTOCs and their dynamics in this system, we will develop some necessary technologies. The main hammer here is to calculate the autocorrelator dynamics.

\subsubsection{Autocorrelator}
In this subsection, we will derive the dynamics of the autocorrelator by expanding to the second order in $\Delta t$ (first order will vanish, as we will see below), and we will specify the variance of the coupling constant $J_A$ in this process. The autocorrelator $G_{W,W}(t)$ of an operator $W$ evaluated at the infinite temperature (denoted by $\expval{...}_{\infty}$) is defined by
\begin{align}
    G_{W,W} (t) \equiv \expval{W(t)W}_\infty = \expval{U^\dagger(t)WU(t)W}_\infty
\end{align}
where $W(t) = U^\dagger(t)WU(t)$ is the Heisenberg picture time evolved operator. In the following notation, we will omit the $\infty$ for simplicity of notation. For a change in time $\D t$, the change in the autocorrelator is
\begin{align}
    \D G_{W,W}(t) = G_{W,W}(t+\D t) - G_{W,W}(t).
\end{align}

Now, we expand the $U$'s in orders of $\D t$ up to $\cO(\D t^2)$ because the first order $H\D t $ is linear in $H$. Terms linear in $H$ will vanish upon taking the ensemble average ($\Expval{\cdot}$) over the Gaussian distribution, due to the linear dependence in the coupling constant $J_{A}$ in each of terms of $H$. The ensemble average ($\sim\Expval{J_A}$) is taken to be 0 here without loss of generality. 

Note that we have not specified the variance of $J_A(t)$ so we need to be cautious as we expand. We denote the Heisenberg picture of $W$ as $W_H\equiv U^\dagger(0,t)WU(0,t)$ and expand $\D G_{W,W}(t)$
\begin{align}
    \D G_{W,W}(t) &= \D t^2 \expval{H W_H H W - \frac{1}{2}H^2W_H W - \frac{1}{2}W_HH^2W} \nonumber\\
    &= \D t^2 \expval{W_H (H WH - \frac{1}{2}WH^2 - \frac{1}{2}H^2W)}.
\end{align}

If $\comm{H}{W}=0$ then this quantity will simply vanish, and the autocorrelator is a constant. However, since $H$ contains every possible 2-Pauli term and $W$ is an arbitrary Pauli string, they generically don't have to commute. We can illuminate the process by taking the ensemble average of $\D G_{W,W}(t)$ in the  $t\rightarrow t+\D t$ instance, recalling that 
\begin{itemize}
    \item $H=\sum_A J_A O_A$;
    \item The variance $\s_J^2$ of the coupling constant $J_A$ is independent of indices $A$;
    \item Ensemble average of terms linear in $J_A$ is zero, $\Expval{J_A} = 0$.
\end{itemize}

Putting them together, we have
\begin{align}
    \Expval{\D G_{W,W}}_{t+\D t} &= \D t^2 \sum_{A,A'} \d_{AA'} \s^2_J \expval{W_H \Big(O_A W O_{A'} - \frac{1}{2}WO_AO_{A'} - \frac{1}{2}O_AO_{A'}W\Big)}\nonumber \\
    &= \D t^2 \s^2_J \sum_A\expval{W_H (O_A W O_A - W)}.
\end{align}

    Consider a simple example where $W$ is a single Pauli, which either commutes or anticommutes with $O_A$. We count the anticommuting terms in $H$ for a non-trivial result. Take $W=\s^z_1$, whose label is chosen without loss of generality, terms in $O_A$ (weight-2) must contain $\s^x_1$ or $\s^y_1$ to anticommute with $W=\s^z_1$ (2 choices), and the other Pauli in that term doesn't matter ($3(N-1)$ redundancy). This concludes our counting and the above equation is now
\begin{align}
    \Expval{\D G_{W,W}}_{t+\D t} = -(2\cross 3 (N-1)) \D t^2 \s_J^2 \expval{2 WW_H}.
\end{align}

Note that $\expval{WW_H}$ is the autocorrelation up to the previous time step $t-\D t$. To demand that $\Expval{\D G_{W,W}}$ change by an order unity amount in unit time $\D t$, we require that 
\begin{align}
    \s^2_J = \frac{\g}{12(N-1)\D t}
\end{align}
where the $\frac{1}{\D t}$ dependence is common in Brownian processes. And we group the rest into $\g$ such that it is the rate of decay of $\Expval{G_{W,W}(t)}$ in the following sense: if we include the ensemble averaging for all the previous time steps $0\rightarrow t$, we can rewrite the change in autocorrelator cleanly as
\begin{align}
    \Expval{\D G_{W,W}} = -\g \D t \Expval{G_{W,W}(t)}.
\end{align}

Solving this gives us the equation of motion for the ensemble-averaged autocorrelation of operator $W$
\begin{eqnarray}
    \Expval{G_{W,W}(t)} &= \expval{W^2}e^{-\g t}
    \label{eq:autocorrelator_app}
\end{eqnarray}
where the initial value $\expval{W^2}$ is the square of the Hilbert-Schmidt norm of the observable $W$, which we set to 1. This determines the physical meaning of the time unit $\gamma^{-1}$: it sets the decay time of autocorrelations of single Pauli operators.

We may also consider $W$ to be a general Pauli string of weight $w$. The counting is slightly different. $O_A = \s_i^\a\s_j^\b$ contains two Pauli's at different sites. At least one should be at the same site but of different Pauli as one of those contained in $W$ for them to possibly \underline{anticommute}. For example, if $W=...\s_1^z...$, then $O_A = \s_i^\a\s_j^\b$ must have one of the two Pauli's being either $\s_1^x$ or $\s_1^z$ for anticommutation to hold. For a given $W$ of weight $w$, there are $2w$ choices for the first Pauli in $O_A$.

There are two cases for the other Pauli in $O_A$, which has to \underline{commute} with the rest of the Pauli's in $W$. It is 
\begin{enumerate}
    \item either on a site not contained by $W$;
    \item or on a site contained by $W$.
\end{enumerate}

In case 1, it can be any Pauli's (e.g. $\s_j^\b\notin W$ and can have any $\b$), and there are $3(N-w)$ choices for this case. In case 2, it has to be the same kind of Pauli as the one in $W$ (e.g. $\s_j^\b\in W$), and there are $w-1$ choices for this case). So all the above counting amounts to an extra factor of $\frac{2w\qty[3(N-w)+(w-1)]}{6(N-1)} $ in front of $\g$ to swap out the counting done for $W$ being a single Pauli (note that the $6(N-1)$ was the counting done for a single Pauli). We get a new decaying dynamics for a generic Pauli string of weight $w$
\begin{align}
    \Expval{G_{W,W}(t)} = \expval{W^2} \exp{-\frac{w\qty[3(N-w)+(w-1)]}{3(N-1)}\g t}\label{eq:counting}
\end{align}
where $w=1$ recovers the single Pauli dynamics that we obtained previously.

\subsubsection{Cross correlator}
Cross correlator is a variation of the autocorrelation, given by 
\begin{align}
    G_{W,V}(t) = \expval{W(t)V}
\end{align}
where $W,V$ are taken to be distinct Pauli string, that is, $G_{W,V}(t=0)=\expval{WV}=0$. We can then follow the same kind of counting in the autocorrelator case and obtain
\begin{align}
    \Expval{G_{W,V}(t)} = \expval{WV} \exp{...} = 0.
\end{align}

So a pair of initially orthogonal operators' time-ordered correlator will remain zero in this fashion for all time.

\subsubsection{Dynamics of the state}
To get the dynamics of the states, we expand the initial state $\r_0$ in the Pauli basis
\begin{align}
    \r_0 = \sum_P a_P(0) P.
\end{align}

The ensemble-averaged dynamics of this state is
\begin{align}
    \Expval{\r(t)} = \Expval{U^\dagger (t) \r_0 U(t)} = \Expval{\sum_Pa_P(t)P}.
\end{align}

Full information of the state dynamics is encoded in the coefficients $a_P(t)$, whose ensemble averaged dynamics is
\begin{align}
    \Expval{a_P(t)} = \Expval{\expval{P\r(t)}}&= \Expval{\expval{PU^\dagger(t)\r_0 U(t)}} \nonumber\\
    &= \Expval{\sum_{Q} a_P(0)\expval{PU^\dagger(t)QU(t)}} \nonumber\\
    &= \sum_P a_P(0)\d_{PQ}e^{-\g_Qt} \nonumber\\
    &= a_Q(0)e^{-\g_Qt}.
\end{align}
where in the second last step we used the results from the cross-correlator and the autocorrelator, and we define $\g_Q\equiv \g w_Q \frac{3(N-w_Q)+(w_Q-1)}{3(N-1)}$. Most of the coefficients decay quickly to 0, except for the identity operator $\bI$, which has weight 0 so the decay rate is $\g_{\bI} = 0$. The initial coefficient for identity is 
\begin{align}
    a_\bI(0) = \frac{1}{2^n}\Tr{\bI \r_0} = \frac{1}{2^N}.
\end{align}

So the \textit{ensemble average} of any initial state $\r_0$ after some long enough time $t \gg 1/\g$ will give us a maximally mixed state. This is to be distinguished from the open quantum system case. Every instance of this ensemble preserves the purity of its initial state. If $\r_0$ is a pure state, then unitary time evolution guarantees that $\r=U\r_0U^\dagger$ is still pure. It is the averaging over all the instances that will give rise to a unique mixed state. 

\subsubsection{OTOCs and the dynamics of operator weight probability}
Now we are ready to work with the Out-of-Time-Order Correlators (OTOCs). Here, we use the squared commutator definition of OTOCs
\begin{align} 
    C_{W,V}(t) \equiv \expval{ [W(t), V]^\dagger [W(t), V] }.
\end{align}

It is related to the other widely used definition of the OTOCs up to a constant shift,
\begin{align}
    C^{(4)}_{W,V}(t) \equiv \expval{W^\dagger(t)V^\dagger W(t)V} = 2 - 2\text{Re}(C_{W,V}).\label{eq:4pt_def_otocs}
\end{align}
They are proposed measures for information scrambling and for quantum chaos for up to the second moment in $U$. Consider the case where $W,V$ are both Pauli strings for simplicity. We can expand the operators in the Pauli basis $W(t) = \sum_P c_P P$ and rewrite the squared commutator
\begin{align}
    C_{W,V}(t) = \sum_{P,P'\text{anticommuting with V}}4c_P c_{P'}\langle VPP'V \rangle_\infty = \sum_{P|\{P,V\}=0}4c_P^2.
\end{align}

Note that if $\comm{P}{P'}=0$ the OTOC would simply vanish. We take the Hilbert-Schmidt norm of  operators  $||O||_2\equiv\sqrt{\tr(O^\dagger O)}$  to be 1. In this sense, if we expand the operator, we get $\tr(O^\dagger O) = \sum_P \abs{c_P}^2 = 1$, which, under unitary evolution, will remain 1. This allows the interpretation of the set of squared coefficient $\{\abs{c_P}^2\}_P$ as probability distributions of the operator over the Pauli basis. With this understanding, OTOCs can be interpreted as  a measure of how the two operators "perceive each other" over time by measuring the total probability of how much they would anticommute. 

In addition, since no direction is special, $\Expval{c_P^2}$ can only depend on the weight of $P$, instead of the specific operator content of it. We can then further group the anticommuting terms based on their weights. It is then convenient to define a total probability $h_w$ for the operator to be at weight $w$ given by
\begin{align}
    h_w \equiv \sum_{P|\text{wt}(P)=w} \mathbb{E} [c_P^2] = 3^w{N\choose w} \Expval{c_P^2} = 3^w{N\choose w} h^{(1)}_w \label{eq:hw_and_hw1}
\end{align}
where we use $h_w^{(1)}$ to denote the ensemble-averaged probability of being on one specific Pauli string at weight $w$. It doesn't matter which one.

From here on, we will assume $V$ as a single Pauli. Taking the ensemble average of the OTOC and counting all the Pauli strings with weight $w$ that anticommute with $V$ (we could imagine $V=\s^z_1$ without loss of generality), this would give us a counting factor of $2\cross 3^{w-1}\cross {N-1\choose w-1}$. Putting this back in, we could write the squared commutator in terms of the total probability that is only dependent on the weight:
\begin{align}
    \mathbb{E}[C_{W,V}(t)] =& \sum_{w|\operatorname{wt}(P)=w\wedge\{P,V\}=0} 4h_w^{(1)}\nonumber\\
    =& \sum_w \frac{4\qty(2\cross 3^{w-1}{N-1\choose w-1})}{3^w {N\choose w}} h_w\nonumber\\
    =&\sum_w \frac{8w}{3N} h_w(t).\label{eq:otoc and h_w}
\end{align}
where we use Eq.~\eqref{eq:hw_and_hw1} for the second equality. In addition, we observe that $\sum_w wh_w(t) = \overline{w}(t)$ is the average weight (averaged over the probability distribution $\{h_{w}\}_w$), so the behavior of the OTOC is governed by this notion of average weight of the operator $W$ under the assumption that $V$ is a single Pauli. As $t\rightarrow\infty$, we expect $\overline{w}(t)\rightarrow \frac{3}{4}n$ because there are 4 total possibilities $\{\s^x, \s^y, \s^z, \bI\}$ and 3 of them contribute to the weight. So at late time, we expect $\mathbb{E}[G_{W,V}(t)]\rightarrow 2$.

Now we have an expression of the OTOC in terms of the dynamics of the probability distribution over weight $\{h_w\}_w$. To obtain dynamics for $h_w$, we can expand $c_P(t)$ for a small change $\D t$. 
\begin{align}
    \D c_P(t) &= c_P(t+\D t) - c_P(t).
\end{align}
With similar reasoning for the expansion down for the autocorrelation, anything linear in $O_A$ will vanish upon taking the trace, and anything linear in $J_A$ will vanish upon taking the ensemble average. So we need to expand to $\cO(\D t^2)$
\begin{align}
    \Expval{\D c_P^2} = - \D t^2 \Biggl(\Bigl\langle PW_H\Bigr\rangle 
 \Expval{\Bigl\langle P\comm{H}{\comm{H}{W_H}}\Bigr\rangle} + \Expval{\Bigl\langle P\comm{H}{W_H}\Bigr\rangle}^2\Biggr).\label{eq:delta c_P^2}
\end{align}

We will deal with the \underline{first term} and the \underline{second term} separately. Note that the \underline{first term} will give us results similar to those derived for the autocorrelators, while the \underline{second term} will give rise to something new. 

We start with the \underline{first term}
\begin{align}
    \Expval{\Bigl\langle P \comm{H}{\comm{H}{W_H}} \Bigr\rangle} = \sum_A\s_J^2 \Bigl\langle 2P(W_H-O_AW_HO_A)\Bigr\rangle.\label{eq:first term}
\end{align}

This term will vanish if $\comm{O_A}{W}=0$, so we count the anticommuting terms in the similar way as done in the autocorrelation case, described above Eq.~\eqref{eq:counting}. This gives us a factor of $2w\Bigl(3(N-w)+(w-1)\Bigr)$
\begin{align}
    \Expval{\Bigl\langle P \comm{H}{\comm{H}{W_H}} \Bigr\rangle} = 8\s_J^2 w\Bigl(3(N-w)+(w-1)\Bigr) \Bigl\langle PW_H\Bigr\rangle.
\end{align}

We now proceed to the \underline{second term}
\begin{align}
    \Expval{\Bigl\langle P\comm{H}{W_H} \Bigr\rangle^2} &= \s_J^2\sum_A\Bigl\langle \comm{P}{O_A}W_H \Bigr\rangle^2 \nonumber\\
    &= 8\s_J^2w\Bigl(3(N-w)+(w-1)\Bigr)    \Bigl\langle PO_AW_H\Bigr\rangle^2 \label{eq:second term}
\end{align}
where the counting follows the same logic as before. The only difference between the first and the second terms is the additional $O_A$. There are two cases where $O_A$ anticommutes with some Pauli string $P$ with weight $w$, and each modifies $w$ differently:
\begin{enumerate}
    \item Case 1, both sites in $O_A=\s^\a_i\s^\b_j$ are present in $P$ ($i,j$ are non-identity for $P$), and one $\s_i^\a$ has same $\a$ label with $i$-th site Pauli in $P$ while the other $\s_j^\b$ has different $\b$ label with $j$-th site Pauli in $P$. In this case $PO_A$ would reduce the weight of $P$ by 1;
    \item Case 2, one site in $O_A=\s^\a_i\s^\b_j$ is present in $P$ and they have different Pauli types. In this case, $PO_A$ would increase the weight of $P$ by 1.
\end{enumerate}

In both cases, we have $PO_A \sim i \times (\text{some Pauli string with weight }w\pm 1)$ so we have an extra overall negative sign. Altogether, the \underline{second term} becomes
\begin{align}
    \Expval{\Bigl\langle P\comm{H}{W_H} \Bigr\rangle^2} &= - 8\s_J^2w  \Bigl(3(N-w)\Expval{c_{w+1}^2}+(w-1)\Expval{c_{w-1}^2}\Bigr) \nonumber\\
    &= - 8\s_J^2w  \Bigl(3(N-w) h_{w+1}^{(1)}+(w-1) h_{w-1}^{(1)}\Bigr)
\end{align}
where $\Expval{\cdot}$ on the RHS comes from ensemble-averaging over all previous time steps. 

Combining the \underline{first term} and the \underline{second term}, we have
\begin{align}
    \D h_w^{(1)} &= 8\s_J^2w\D t^2 \Bigl[ 
     (w-1)h_{w-1}^{(1)} 
    - \bigl(3(N-w)+(w-1)\bigr)h^{(1)}_w 
    + 3(N-w)h_{w+1}^{(1)}\Bigr].
\end{align}

Recall that $\g=12(N-1)\s_J^2\D t$, and take $\D t\rightarrow 0$, we have
\begin{align}
    \frac{1}{\g}\dv{h_w^{(1)}}{t}  =  
    &- \frac{2w\bigl(3(N-w)+(w-1)\bigr)}{3(N-1)}h^{(1)}_w +\frac{2w(w-1)}{3(N-1)} h_{w-1}^{(1)} + \frac{2w(N-w)}{(N-1)}h_{w+1}^{(1)}.
\end{align}

We could massage this a bit more with the total probability $h_w={n\choose w} 3^w h_w^{(1)}$. 
\begin{align}
    \frac{1}{\g}\dv{h_w}{t} = & -\frac{2w \Bigl((w-1)+3(N-w)\Bigr)}{3(N-1)} h_w + \frac{2(N-w+1)(w-1)}{ (N-1)}h_{w-1} + \frac{2w(w+1)}{3(N-1)}h_{w+1}.\label{eq:h_w master dyanmics}
\end{align}

The $h_w$ on the RHS contributes to the decay of total probability at a given weight, while the $h_{w+1}$ and $h_{w-1}$ terms on the RHS are inflowing of probability contributed by the nearby weights. This is the master equation for the dynamics of the total probability of different weight components of a given operator $W(t)$, and it governs the dynamics of the OTOCs through Eq.~\eqref{eq:otoc and h_w}.

\subsubsection{Dynamics of the OTOCs in the dilute limit $\overline{w}\ll n$}
We reproduce Eq.~\eqref{eq:otoc and h_w} below, and recall that we assumed $V$ is a single Pauli for simplicity
\begin{align}
    \mathbb{E}[C_{W,V}(t)] =\sum_w \frac{8w}{3N} h_w(t)\sim  \overline{w}.
\end{align}

Consider a dilute limit $\overline{w}\ll N$, or take the system size to infinity $N\rightarrow\infty$, then we could simplify Eq.~\eqref{eq:h_w master dyanmics}
\begin{align}
    \frac{1}{\g}\dv{\overline{w}}{t} &\approx -\sum_w 2w^2 h_w + 2w(w-1)h_{w-1} \nonumber\\
    &\approx -\sum_w 2w^2 h_w + \sum_w 2(w+1)w h_w \nonumber\\
    &= 2\overline{w} + \cO\qty(\frac{w}{N}).
\end{align}

This gives $\overline{w}(t) = \overline{w}(0)e^{2\g t}$ the exponential operator growth in early time. At late time, the average weight grows to $\frac{3}{4}n$, which gives us the maximum value of $\Expval{C_{W,V}(t)}=2$. This can be more illuminating if we consider the other definition of the OTOCs $C^{(4)}_{W,V}(t)\equiv \expval{VW(t)VW(t)}_ = 2 - C_{W,V}(t)$. So (under ensemble averaging for our case), having $C_{W,V}\rightarrow 2$ means, at late time, information in the system is fully scrambled, the correlations between operators decay to zero, and $C^{(4)}_{W,V}(t) \rightarrow \expval{VW(t)}^2 \rightarrow 0$. 

\subsection{Dressed OTOC and echo in the discrete Brownian cluster model\label{app:dressed_otoc}}
In this section, we generalize the operator dynamics to the dressed OTOC, defined to be a variation of the squared commutator definition of the OTOCs
\begin{align}
    \dress{C}_{W,V} = \expval{\comm{\dress{W}(t)}{V}^\dagger \comm\Big{W(t)}{V}}.
\end{align}

The only difference here is that we now include a perturbed Hamiltonian in one of the time evolution branches. For notation, we use $\dress{\cdot}$ to denote any quantity that is ``dressed", put simply, any quantity that is with a perturbation in the backward branch of time time evolution. Here, $\dress{W}(t)$ is the Heisenberg picture of an operator $W$ evolved by a perturbed Hamiltonian
\begin{align}
    \dress{H} &= \frac{1}{S}\bigg((1-p)\sum_A J_A O_A + p\sum X_A O_A\bigg) \nonumber\\
    &= \frac{1}{S}\sum_A \bigg((1-p)J_A + p X_A\bigg) O_A \nonumber\\
    &= \sum_A \dress{J}_A O_A
\end{align}
where $\dress{J}_A \equiv \big((1-p)J_A + pX_A\big)/S$ is the effective coupling constant for the backward branch as described in Eq.~\eqref{eq:forw_back_covar}. $X_A, J_A$ are independently sampled from the same Gaussian distribution. The overall normalization factor $S$ will be fixed later when we discuss the variance of the perturbed branch. Expanding the dressed OTOCs
\begin{align}
    \dress{C}_{W,V} = \expval{2\dress{W}(t)W(t) - 2\dress{W}(t)VW(t)V}.\label{eq:dressed OTOC square commutator and 4-pt}
\end{align}

The second term is the dressed version of the $C^{(4)}_{W,V}$ described in Eq.~\eqref{eq:4pt_def_otocs}. Unlike in the ordinary OTOCs case, the first term is not a constant, so the maximum value of $\dress{C}_{W,V}(t)$ decays over time due to decaying fidelity $\expval{\dress{W}(t)W(t)}$, also referred to as the echo signal. To properly compare the dynamics of OTOCs and the dressed OTOCs, we divide out the echo signal from the dressed OTOC.

Expand $W,\dress{W}$ in Pauli basis, $\dress{W}(t) = \sum_P \dress{c}_P P$ and $W(t) = \sum_P c_P P$, the second term in the above equation is
\begin{align}
    \expval{\dress{W}(t)VW(t)V} = \sum_{P,P'} \dress{c}_P(t)c_{P'}(t)\expval{PVP'V}.
\end{align}

Similar to the treatment in the ordinary OTOCs case, we consider the anticommuting terms for nonvanishing results
\begin{align}
    \dress{C}_{W,V} &= \sum_{P,P'|\{P,V\}=\{P',V\}=0} 4 c_P \dress{c_{P'}} \expval{P'P}= \sum_{P|\{P,V\}=0} 4 c_P \dress{c}_P
\end{align}
where we suppressed the time dependence notation. The result only depends on weight of the operator, so we further group the $P$'s that anticommute with $V$ according to their weights $w$.

We define a total overlap between the two branches at a given weight $w$, similar to the total probability defined in the ordinary OTOCs case
\begin{align}
    b_w\equiv\sum_{P|wt(P)=w} \Expval{\dress{c}_Pc_P}.
\end{align}

For the overlap of the two branches at one instance of weight $w$ operator, we denote it $b_w^{(1)}\equiv \Expval{\dress{c}_Pc_P}$. The relation between $b_w^{(1)}$ and the overlap at all the operators of weight $w$ is a result of counting
\begin{align}
    b_w = 3^w {N\choose w}b_w^{(1)} \eval_{P|wt(P)=w}.
\end{align}

Exactly which $P$ is chosen for $b^{(1)}_w$ does not matter under ensemble averaging as long as it is weight $w$. Assuming $V$ is a weight one Pauli. Combining the counting for the amount of weight $w$ Paulis that anticommutes with $V$, we have 
\begin{align}
    \Expval{\dress{C}_{W,V}(t)} = \sum_w \frac{4\qty(2\cross 3^{w-1}{N-1\choose w-1})}{3^w {N\choose w}} b_w=\sum_w \frac{8w}{3N}b_w(t) \label{eq:dotoc_bw_relation}
\end{align}

\subsubsection{Relation between perturbation strength $p$, correlation coefficient $r$, and variance $\s_J^2$}
Variance for the coupling constant $J_A$ is $\sigma_J^2$ in the unperturbed Hamiltonian. The coupling constant of the perturbed Hamiltonian should have the same variance so both Hamiltonian wander away (in a Brownian motion sense) at the same rate and we are comparing the operator structure for the two branches on the equal footing. Take $\dress{J}_A \rightarrow \dress{J}_A/S$ for some undetermined variable $S$. For the perturbed case, the variance is
\begin{align}
    \Var{\frac{\dress{J}_A}{S}} = \frac{\Expval{\dress{J}_A^2}}{S^2}&= \Expval{(1-p)^2 J_A^2 + 2p(1-p) J_A X_A + p^2 X_A^2}/S^2 \nonumber\\
    &= \frac{1-2p+2p^2}{S^2}\sigma_J^2
\end{align}

In the last equality, we used the fact that $J_A$ and $X_A$ are independent, $\Expval{X_AJ_A} = 0$, and they have the same variances. So we fix $S=\sqrt{1-2p+2p^2}$ and the perturbed Hamiltonian is modified to
\begin{align}
    \dress{H} = \frac{1}{S}\sum_A \dress{J}_A O_A
\end{align}

The correlation coefficient $r$ for the two branches is
\begin{align}
    r = \frac{\text{Cov}\qty[\frac{\dress{J}_A}{S}, J_A]}{\sqrt{\text{Var}\qty[\frac{\dress{J}_A}{S}]\text{Var}[J_A]}} &= \frac{\mathbb{E}[\dress{J}_AJ_A/S] - \mathbb{E}[\dress{J}_A/S]\mathbb{E}[J_A]}{\s_J^2} = \frac{1}{S} \frac{\mathbb{E}[\dress{J}_AJ_A]}{\s_J^2} = \frac{1-p}{S} \label{eq:r_p_relation}
\end{align}
where in the last equality we recall $\dress{J}_A = (1-p)J_A + pX_A$. In this sense, we tune the correlation between forward and backward time evolution through the perturbation strength $p$, consistent with what the original nuclear spin experiment\cite{Alvarez2015} did. Rearranging the second line gives us Eq.~\eqref{eq:forw_back_covar}
\begin{align}
    \s_J^2 r=r\Expval{J_AJ_A} = \Expval{\dress{J}_AJ_A}
\end{align}
up to a delta function that can be put in by hand.

\subsubsection{Dressed OTOC dynamics}
To solve for the dynamics for the dressed OTOCs, we focus on the dynamics of $b_w$. We start by recalling the following properties
\begin{align}
    \Expval{J_AJ_{A'}} &= \d_{AA'}\s_J^2 \\
    \Expval{X_AJ_{A'}} &= \Expval{J_AX_{A'}}=0\\
    \Expval{X_A\dress{J}_{A'}} &= \d_{AA'}\s_J^2.
\end{align}

Then we work out the dynamics of $b_w^{(1)} = c_P\dress{c}_P$. Note that the choice of $P$ eventually won't matter upon ensemble averaging, but we will use $P$ as a label to remind ourselves that we are dealing with the overlap between two branches at one specific operator $P$ right now.  Expand $\Expval{\D b_w^{(1)}}$ for a change in time $\D t$ up to the second order.
\begin{align}
    \Expval{\D b_w^{(1)}} =& \Expval{b_w^{(1)}(t+\D t) -  b_w^{(1)}(t)} \nonumber\\
    =&-\frac{1}{2} \D t^2 \bE\bigg[ 
    \expval{P\dress{W}(t)}\expval{P\comm{H}{\comm{H}{W(t)}}} + \expval{PW(t)}\expval{P\comm{\dress{H}}{\comm{\dress{H}}{\dress{W}(t)}}} \nonumber \\ 
    &+ 2\expval{P\comm{\dress{H}}{\dress{W}(t)}}\expval{P\comm{H}{W(t)}}\bigg] \label{eq:delta b_w expansion}.
\end{align}
We will work out the three terms separately. The \underline{first term} gives us, upon ensemble averaging,
\begin{align}
    \Expval{\expval{P\dress{W}(t)}\expval{P\comm{H}{\comm{H}{W(t)}}}}&= 2\s_J^2 \Expval{\dress{c}_P  \sum_A \expval{P(\dress{W}-O_A\dress{W}O_A)}}\nonumber\\
    &= 2\s_J^2 \Expval{\dress{c}_P  \sum_{A|\{O_A,P\}=0} 2\expval{P\dress{W}}}\nonumber\\
    &= 8w\sigma_J^2\big[3(N-w) + (w-1)\big]b_w^{(1)} \nonumber\\
    &= 2D_w b_w^{(1)}
\end{align}
where we define $D_w \equiv 4w \sigma_J^2\big[3(N-w) + (w-1)\big]$ for simplicity of notation, and $w\equiv\operatorname{wt}(P)$. The coupling constants $J_A$ and $\dress{J}_A$ in $H$ and $\dress{H}$ have the same variance, the \underline{second term} will yield the same result as the first
\begin{align}
    \Expval{\expval{PW(t)}\expval{P\comm{\dress{H}}{\comm{\dress{H}}{\dress{W}(t)}}}}= \Expval{\expval{P\dress{W}(t)}\expval{P\comm{H}{\comm{H}{W(t)}}}} = 2D_w b_w^{(1)}
\end{align}

From the first two terms, we see that the probability of walking back to the same weight after $\D t$ is the same for both branches. 

For the \underline{third term}, expand the terms in the Hamiltonians
\begin{align}
    \Expval{\expval{P\comm{H}{W(t)}}\expval{P\comm{\dress{H}}{\dress{W}(t)}}} &= \bE\bigg[\sum_{A,A'} \frac{1}{S}\qty[(1-p)J_{A'} + p X_{A'}]J_A  \Bigl\langle\comm{P}{O_A}W(t)\Bigr\rangle \expval{\comm{P}{O_{A'}}\dress{W}(t)} \bigg] 
\end{align}

Here, we have two parts to consider, the coupling constant part and the operator part, the first one gives the proper variance, and the second gives the proper counting factor. First, consider the ensemble average over the coupling constant
\begin{align}
     \bE \qty[\qty((1-p)J_{A'} + p X_{A'})J_A] = \d_{AA'} (1-p) \s_J^2
\end{align}

The operator part follows the counting described in the ordinary OTOCs case (see discussion below Eq.~\eqref{eq:second term},) which gives rise to a $b_{w-1}^{(1)}$ term and a $b_{w+1}^{(1)}$ term. After taking the ensemble average, besides having an additional factor of $r=\frac{1-P}{S}$ from this \underline{third term}, the rest is the same as the ordinary OTOC. As a result, we have a similar master equation of dynamics for the dressed OTOCs
\begin{align}
    \frac{1}{\g}\dv{b_w}{t} = & -\frac{2w \Bigl((w-1)+3(N-w)\Bigr)}{3(N-1)} b_w + r\qty[\frac{2(N-w+1)(w-1)}{ (N-1)}b_{w-1} + \frac{2w(w+1)}{3(N-1)}b_{w+1}]\label{eq:b_w master dyanmics}
\end{align}

The total probability is not conserved, $\sum_{w'} b_{w'}(t) \leq 0$ due to the decaying echo signal as discussed at the beginning of this section in Eq.~\eqref{eq:dressed OTOC square commutator and 4-pt}. To properly account for the dynamics of the dressed OTOCs in Eq.~\eqref{eq:dotoc_bw_relation} and compare it with the ordinary OTOCs, we divide out the echo $\expval{W(t)\dress{W}(t)} = \sum_w b_w$. The renormalized  $c_w\equiv b_w/\sum_w b_w$ will therefore be a proper probability distribution over weight $w$.

In the dilute limit $w\ll N$, which also corresponds to the early time growth, and take $\gamma=1$ for simplicity, the above equation simplifies to
\begin{align}
    \dv{b_w}{t} = -2w b_w + 2r(w-1)b_{w-1}
\end{align}

The probability mass moves towards the larger weight. The competing effects are between the on-site decay for each weight and the probability mass moving from the lower weight. The dressed OTOCs dynamics is described by the dynamics of the average weight $\overline w = \sum_ww b_w$, or the weight averaged over the overlap between the forward and backward branches
\begin{align}
    \dv{\overline{w}}{t} \approx & \sum_w-2w^2b_w + 2rw(w-1)b_{w-1} \\
    = & 2r\overline{w} + 2(r-1)\overline{w^2}
\end{align}
Compared to the ordinary OTOCs, we have extra damping from the higher moment term. We will give more details about the dilute limit in the next appendix.

\subsection{Noisy channel effects}

Finally, to include the effects of decoherence, we must include one more term into Eq.~\eqref{eq:b_w master dyanmics} arising from the channel acting during the forward and backward evolutions. From Eq.~\eqref{eq:op_decoherence} we see that the operator amplitude for an operator of weight $w$ is damped at rate $ \kappa w$. Since $b_w$ is the sum of squares of operator amplitudes of weight $w$, it will be damped at rate $ 2 \kappa w$. Hence, we must include the extra term $-2 \kappa w b_w$ into the right hand side of Eq.~\eqref{eq:b_w master dyanmics}. This yields the full dynamics reported Eq.~\eqref{eq:b_w_eom} in the main text.
\section{Master equation in the dilute limit}
\label{app:master_dilute}

When the probability mass is predominantly on low-weight operators, the dynamics are in the dilute limit, $w \ll N$. The rate matrix in the master equation simplifies to
\begin{equation}
    M_{w,w'} = - 2 w (1+\kappa) \delta_{w,w'} + 2 r (w-1) \delta_{w-1,w'}.
\end{equation}
For simplicity, we solve the case with $\kappa=0$ explicitly here; the full $\kappa$ dependence is obtained by sending $t \to (1+\kappa)t$ and $r \to r_{\text{eff}}$ in the final results.

In this limit, we can construct the exact solution as follows. $M$ is not Hermitian, but it does have a set of eigenvectors which can be used to solve for the dynamics. From its lower triangular form, we immediately obtain the characteristic polynomial of $M$ as
\begin{equation}
    \det(\lambda-M) = (\lambda+2)(\lambda +4) \cdots = \prod_{\ell=1}^\infty (\lambda + 2\ell).
\end{equation}
Hence, the eigenvalues of $M$ are $-2\ell$ for $\ell=1,2,\cdots$. 

The corresponding eigenvectors can be built starting from the $-2$ eigenvector. A short calculation reveals that this eigenvector takes the form of an exponential distribution. Explicitly, the vector
\begin{equation}
    v_1 = \begin{bmatrix}
        1 \\ r \\ r^2 \\ r^3 \\ \vdots
    \end{bmatrix}
\end{equation}
is an eigenvector of $M$ with eigenvalue $-2$. This eigenvector controls the late-time limit we identified in the main text because it is the slowest decaying eigenvector of $M$. Note that it is not normalized as written here, and it can only be normalized when $r<1$. We will comment on other steady states below.

From $v_1$ we obtain a tower of eigenvectors,
\begin{equation}
    v_\ell = \partial_r^\ell v_1,
\end{equation}
including
\begin{equation}
    v_2 = \begin{bmatrix} 0 \\ 1 \\ 2r \\ 3r^2 \\ \vdots
    \end{bmatrix},
\end{equation}
\begin{equation}
    v_3 = \begin{bmatrix} 0 \\ 0 \\ 2 \\ 6r \\ \vdots
    \end{bmatrix},
\end{equation}
and so on. These have eigenvalues $\lambda_\ell = -2\ell$, which includes the $\ell=1$ case $v_1$.

The time evolution of $v_\ell$ under the $M$ dynamics is simply $e^{\lambda_\ell t}v_\ell$, so if the initial distribution can be expanded in terms of $v_\ell$, we have a simple rule for its subsequent time evolution. 

If the initial state is
\begin{equation}
    b(0)=\begin{bmatrix}
        1 \\ 0 \\ \vdots
    \end{bmatrix},
\end{equation}
then a short calculation shows that
\begin{equation}
    b(0) = v_1 + (-r) v_2 + \frac{(-r)^2}{2} v_3 + \cdots.
\end{equation}
The coefficient of $v_\ell$ is the $\ell-1$-th term in the Taylor series for $\exp(-r)$. Hence, we obtain the dynamical evolution
\begin{equation}
    b(t)  = e^{-2 t} v_1 + e^{-4t} (-r) v_2 + e^{-6t} \frac{(-r)^2}{2} v_3 + \cdots.
\end{equation}

Various useful properties can then be computed. First, using the vector $u$ of all ones, the total probability mass is expressed as $u^T b$. Using $u^T v_1 = (1-r)^{-1}$, we obtain
\begin{equation}
    u^T b(t) = e^{-2t} \sum_{\ell=0}^\infty \frac{(-r e^{-2t})^\ell}{\ell!} \partial_r^\ell \frac{1}{1-r} =  \frac{e^{-2t}}{1-r+r e^{-2t}}.
\end{equation}
The normalized probability distribution is
\begin{equation}
    c(t) = \frac{b(t)}{u^T b(t)}.
\end{equation}

Similarly, the average weight can be obtained using the weight vector $\check{w} = \begin{bmatrix}
    1 & 2 & 3 & \cdots
\end{bmatrix}^T$. Another calculation gives 
\begin{equation}
    \langle w \rangle_c = \frac{\dress{w}^T b(t)}{u^T b(t)} = \frac{1}{1-r+r e^{-2t}}.
\end{equation}
We can now include the effects of non-zero $\kappa$ easily. Recalling that $r_{\text{eff}}=\frac{r}{1+\kappa}$ and the rescaling of time $t\rightarrow (1+\k)t$ the average weight as a function of time is
\begin{equation}
    \langle w \rangle_c = \frac{1}{1-r_{\text{eff}}+r_{\text{eff}} e^{-2(1+\kappa)t}}.
    \label{eq:w_general_result}
\end{equation}
This is the exact solution for the average weight in the dilute limit when the initial condition is concentrated on weight one operators, as can be confirmed via numerical integration of the master equation. 

\subsection{Other initial conditions\label{app:w0_dilute_dynamics}}
Consider the case when the initial condition is concentrated on operators of weight $w_0$. Starting with the master equation in the dilute limit 
\begin{equation}
    \dv{c_w}{t} = -2wc_w + 2r(w-1)c_{w-1}+2(1-r)\expval{w}_c
\end{equation}

Define generating function $C(z,t)\equiv \sum_w c_w z^w$ whose time derivative is 
\begin{equation}
    \pdv{C}{t} = -2z\pdv{C}{z}+2rz^2\pdv{C}{z} + 2(1-r)\expval{w}_c C
\end{equation}

Define a new variable $\tilde C=C\exp{-\int_0^t 2(1-r)\expval{w}_c(\tau) \dd \tau}$ to simplify the above equation. The resulting equation for $\tilde C$ is
\begin{equation}
    \pdv{\tilde C}{t} = -2z\pdv{\tilde C}{z}+2rz^2\pdv{\tilde C}{z}
\end{equation}
This first-order PDE can be solved using the method of characteristics. Along characteristic curves where $\dv{z}{t}=-2z+2rz^2$, we have $\dv{\tilde C}{t}=0$. The characteristic curves can be found by solving the differential equation for $z(t)$, and the solution is then
\begin{equation}
    z(t) = \frac{1}{1/z_0 \cdot e^{2t} - r(e^{2t} - 1)}
\end{equation}
where $z_0 = z(0)$ is the initial value of $z$. Since $\tilde{C}$ is constant along characteristics, we have $\tilde{C}(z(t),t) = \tilde{C}(z_0,0)$.

For an initial condition concentrated at weight $w_0$, we have $C(z,0) = z^{w_0}$, so $\tilde{C}(z,0) = z^{w_0}$. Therefore, the solution is (see $r  = 1$ version in \cite{zhou_operator_2019})
\begin{equation}
    \tilde{C}(z,t) = z_0^{w_0} = \left(\frac{z}{1 - rz(1-e^{-2t})}\right)^{w_0}
\end{equation}

To recover $C(z,t)$, we need to multiply by $\exp{\int_0^t 2(1-r)\expval{w}_c(\tau) \dd \tau}$. This factor depends on the average weight, which itself depends on the solution. To find the average weight, we use $\expval{w}_c = \frac{\partial_z C(z,t)|_{z=1}}{C(1,t)}$. After some calculation, we obtain
\begin{equation}
    \expval{w}_c = \frac{w_0}{1-r+re^{-2t}}
\end{equation}

This generalizes our earlier result for $w_0=1$. Including the effects of non-zero $\kappa$, we have
\begin{equation}
    \expval{w}_c = \frac{w_0}{1-r_{\text{eff}}+r_{\text{eff}}e^{-2(1+\kappa)t}}
\end{equation}

The probability distribution itself can be recovered from the generating function by taking derivatives:
\begin{equation}
    c_w(t) = \frac{1}{w!}\left.\frac{\partial^w C(z,t)}{\partial z^w}\right|_{z=0}
\end{equation}


\subsection{Finite time crossovers}
\label{app:dilute_ftc}

\begin{figure}
        \centering
    \includegraphics[width=0.8\linewidth]{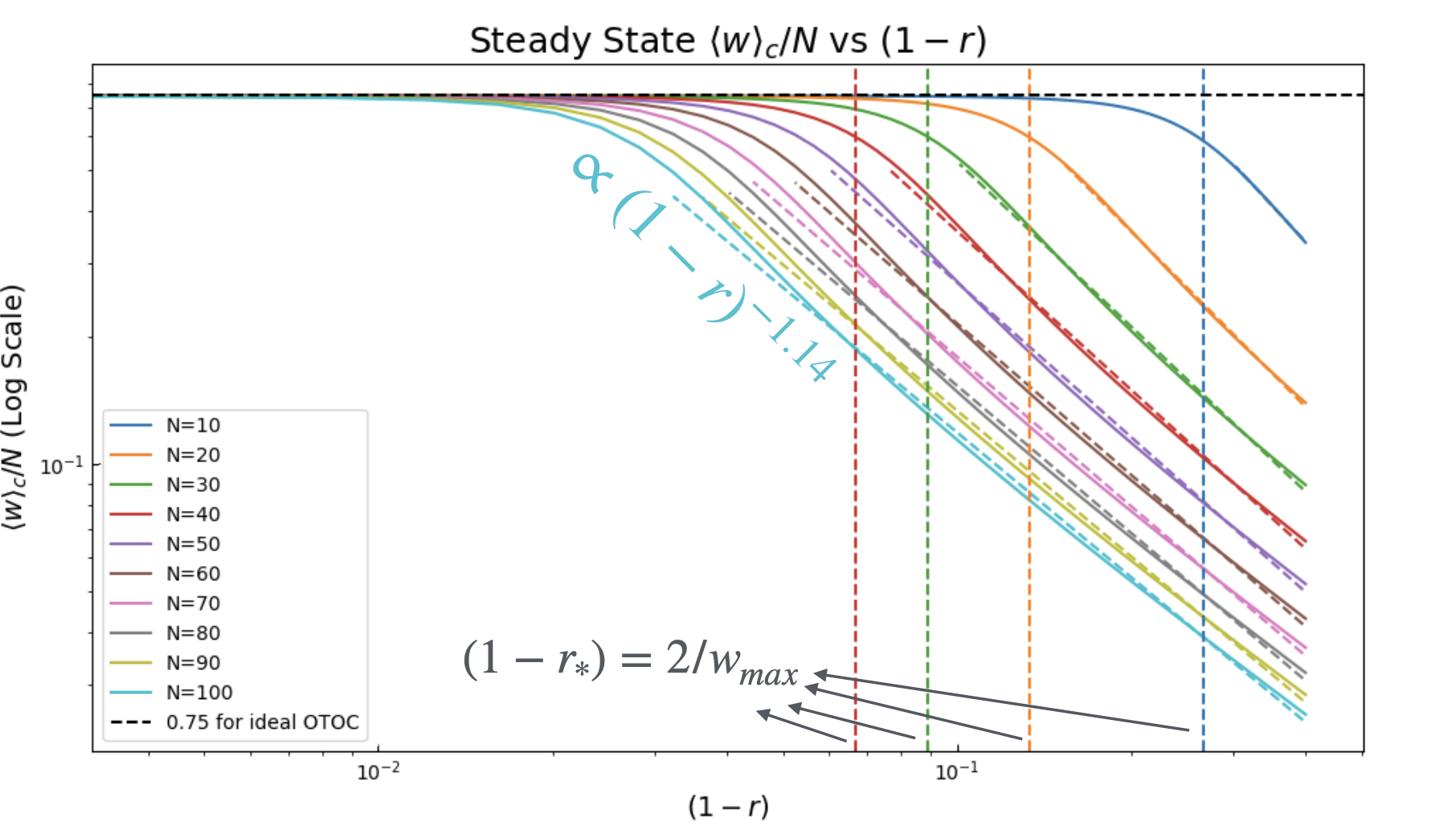}
    \caption{At late time, both the ideal OTOCs and the ROTOCs reach a maximum value. Their dynamics are described by Eq.~\eqref{eq:rotoc_avg_weight}. The saturation values normalized by $N$ for $\expval{w}_c$ against $(1-r)$ are plotted here for various system sizes $N$ and initial condition $w_0=1$. Since we are only concerned with the saturation values, we set $\k=0$ according to the discussion surrounding Eq.~\eqref{eq:r_eff}-\eqref{eq:bw_kappa=0}. For the ideal OTOCs, $\expval{w}_c^{\text{max}} = \frac{4}{3N}$ which gives $C_{W,V}^{\text{max}}(t)=2$. The ROTOCs remain close to the ideal OTOCs when $1-r< 2/w_{\text{max}}$ according to the finite size crossover analysis following Eq.~\eqref{eq:inflection_point}. For reference, four crossover points are plotted in dashed vertical lines, corresponding to $N=40,30,20,10$ in that order. When $1-r>2/w_{\text{max}}$, this is equivalent to the late time dynamics. The slopes fitted to this region gives $\expval{w}_c/N\propto (1-r)^{-\alpha}$ for $\alpha=1.48, 1.38, 1.30, 1.25, 1.22, 1.18, 1.16, 1.15, 1.14$ corresponding to $N=10\cdots 100$ in that order, where $\alpha \rightarrow 1$ for increasing system size, recovering Eq.~\eqref{eq:late_time_w_c} in the $N\rightarrow\infty$ limit.}
    \label{fig:saturation} 
\end{figure}

Here we record some formulas for the finite time crossover analysis. Setting $\kappa=0$, we write the average weight as a function of $g = r(1-e^{-2t})$ as
\begin{equation}
    \langle w \rangle_c = \frac{1}{1-g}.
\end{equation}
The crossover point, $r_*$, is defined such that 
\begin{equation}
    \partial_t^2 \langle w \rangle^{r=r_*}_c (T) = 0,
\end{equation}
where $T$ is the maximum time. This condition is
\begin{equation}
    \left[(1-g) \partial_t^2 g + 2 (\partial_t g)^2 \right]_{r=r_*,t=T} = 0,
\end{equation}
where $\partial_t g= 2 r e^{-2t}$ and $\partial_t^2 g = - 4 r^2 e^{-2t}$. 

The crossover point can be obtained analytically in the large $T$ limit where $r_*$ is close to $1$. Using $g(T) \approx r$, we obtain the condition
\begin{equation}
    (1-r_*) 4 r_*^2 e^{-2T} = 8 r_*^2 e^{-4T} \to (1-r_*) = 2 e^{-2T}.
\end{equation}
This gives
\begin{equation}
    r_* = 1 - 2 e^{-2T} + \cdots = 1 - \frac{2}{w_{\max}} + \cdots
\end{equation}
where again the maximum average weight is $w_{\max} = e^{2T}$, see Fig.~\ref{fig:saturation} for an illustration, where the steady state value of $\langle w \rangle_c /N$ depends on $(1-r)$, and we generally expect a crossover from $3N/4$  to $\frac{1}{1-r}$ at $1-r_* \approx 2/w_{\text{max}}\propto 1/N$.

In the main text, we choose $r_\text{crit}=r_*$ to plot the scaling collapse in Fig.~\ref{fig:scaling_0.9956}. Here, we provide two additional calculations in Fig.~\ref{fig:scaling_collapse} for two arbitrarily picked $r_\text{crit}$ values, where we observe that the time evolution of $\expval{w}_c$ still collapses into two branches, indicating an absence of critical behavior in this system.

\begin{figure}[t]
    \centering
    \includegraphics[width=0.48\textwidth]{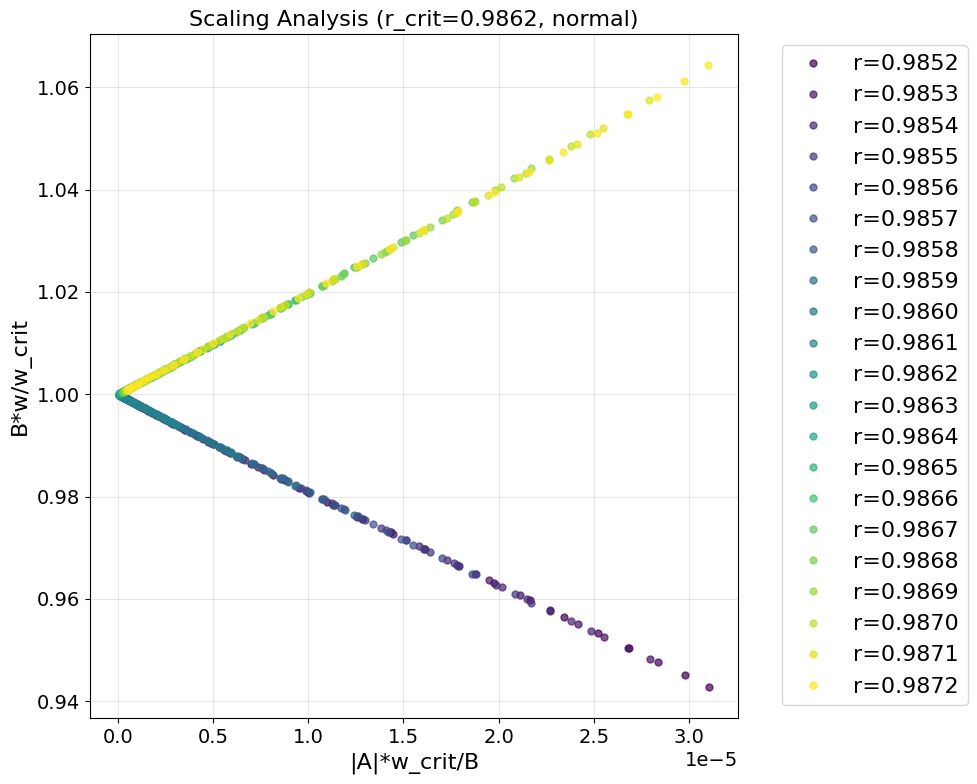}
    \hspace{0.02\textwidth}
    \includegraphics[width=0.48\textwidth]{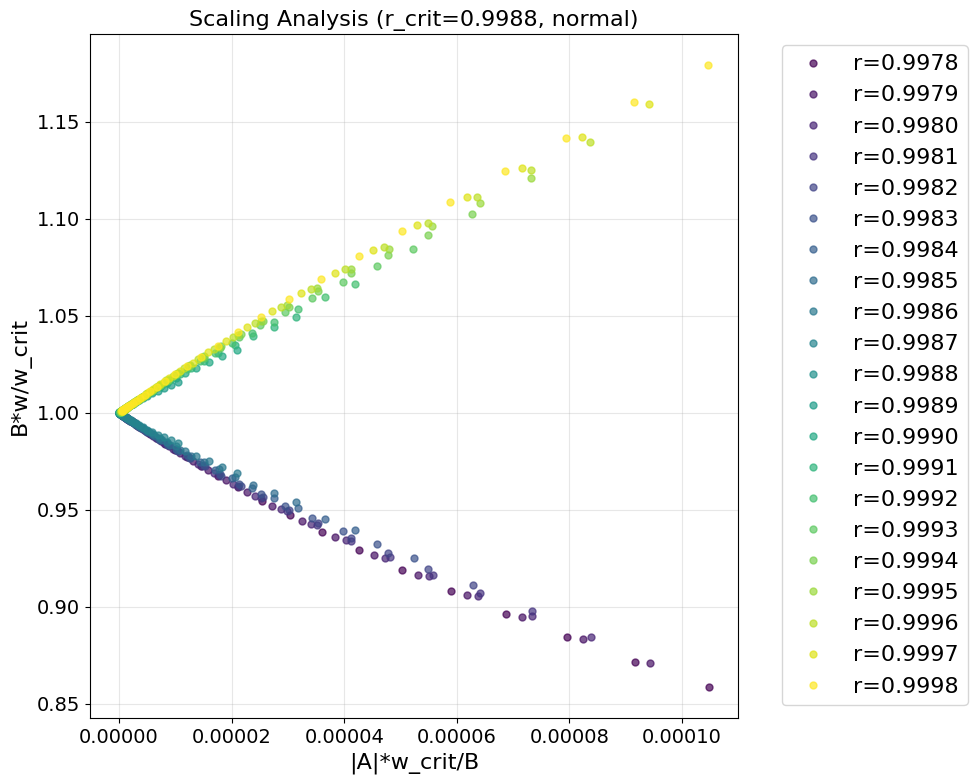}
    \caption{The critical parameter $r_{\text{crit}} = 0.9862$ (left) and $r_{\text{crit}} = 0.9988$ (right) are chosen arbitrarily, and we observe the same kind of collapse into two branches as in Fig.~\ref{fig:scaling_0.9956}. The cutoff time $T$ is determined similarly as in Fig.~\ref{fig:scaling_0.9956}.}
    \label{fig:scaling_collapse}
\end{figure}

\begin{figure}
    \centering
    \includegraphics[width=0.7\linewidth]{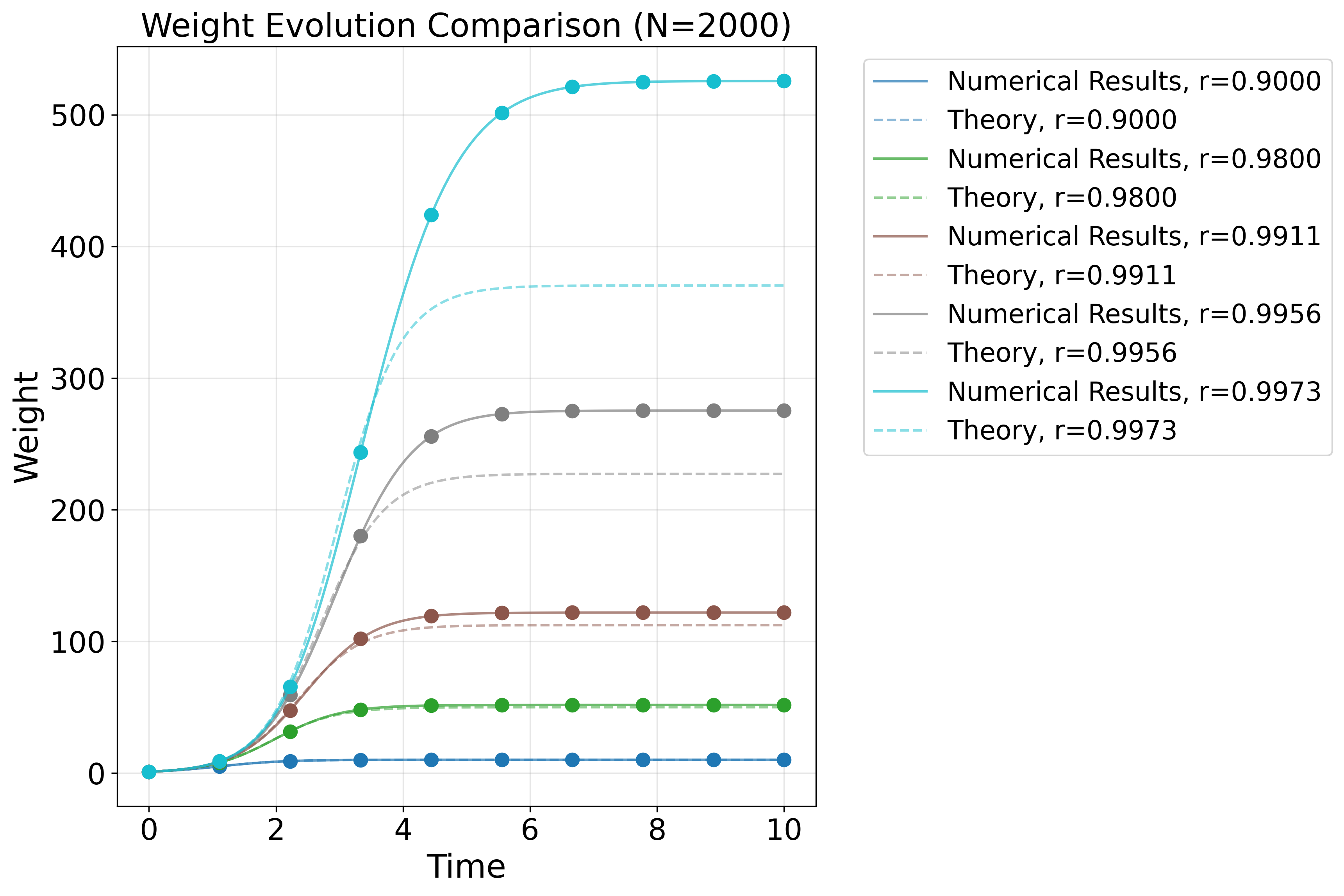}
    \caption{Time evolution of average weight $\langle w \rangle_c$ for $N=2000$ with various $r$ values. The critical value $r_*=0.9956$ (determined by finding when $\partial_t^2\langle w \rangle_c (r=1, t)$ reaches its maximum) separates curves that are beginning to saturate ($r>r_*$) from those still in the growth phase ($r<r_*$). For $r>r_*$, numerical results deviate from dilute limit predictions due to edge effects, while for $r<r_*$, the numerical results closely follow the dilute limit approximation. This demonstrates the equivalence between the dilute limit and the small $r$ value regime.}
    \label{fig:r_dilute_equivalence}
\end{figure}
\section{Numerics for the Full Dynamics\label{app:numerics}}

\begin{figure}[t]
    \centering
    \begin{tabular}{cc}
        \includegraphics[width=0.48\textwidth]{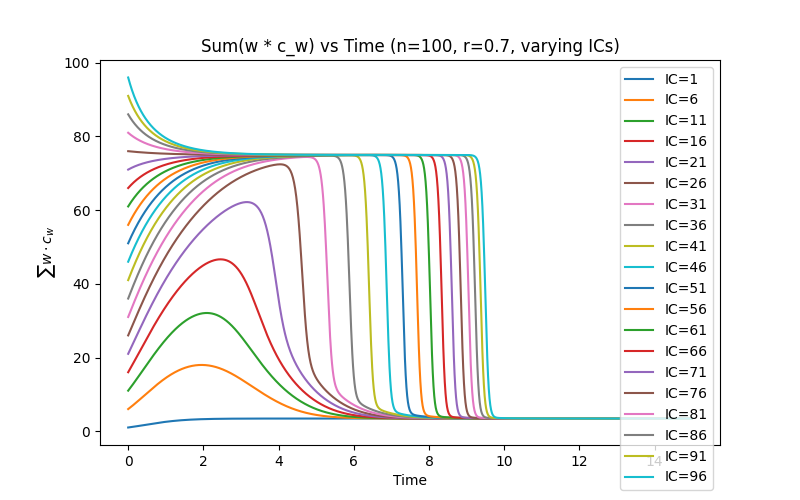} & 
        \includegraphics[width=0.48\textwidth]{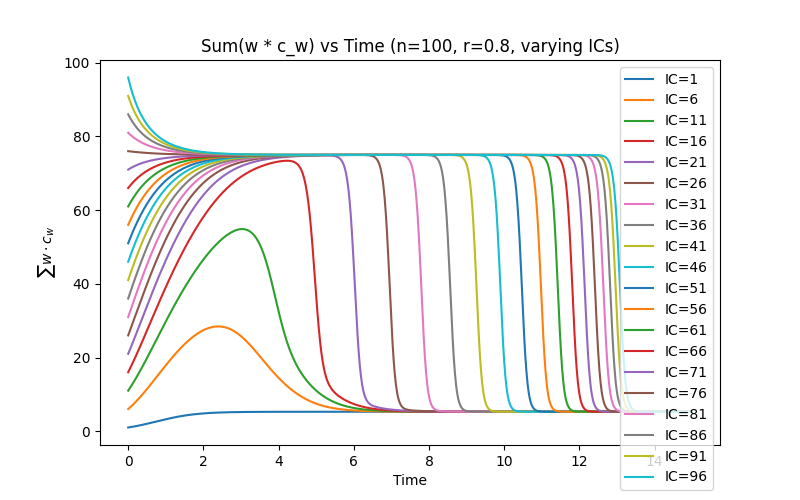} \\
        (a) & (b) \\
        \includegraphics[width=0.48\textwidth]{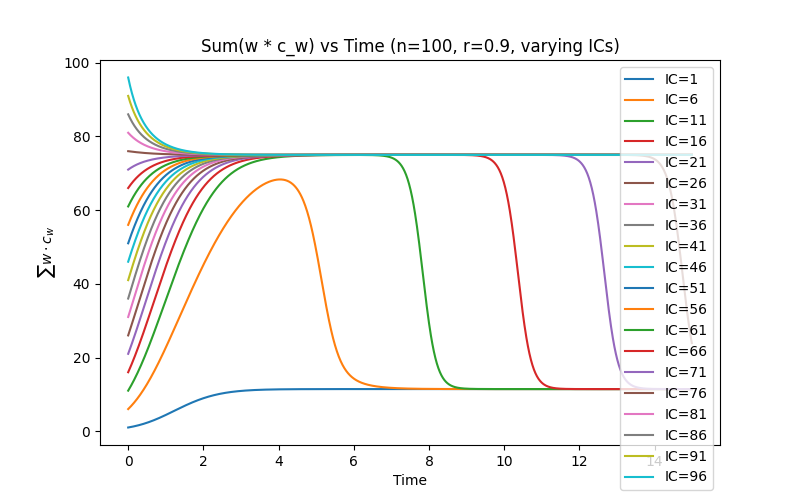} &
        \includegraphics[width=0.48\textwidth]{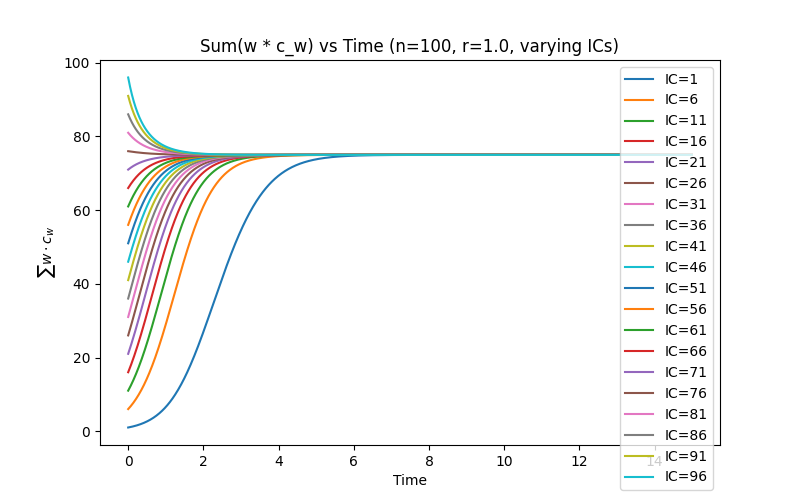} \\
        (c) & (d)
    \end{tabular}
    \caption{Evolution of the ROTOC (average weight $\langle w \rangle_c$) for a system with $N=100$ qubits over 15 time units with varying initial conditions. Each panel shows results for a different correlation value: (a) $r=0.7$, (b) $r=0.8$, (c) $r=0.9$, and (d) $r=1.0$ (ideal OTOC case). All initial conditions eventually converge to the same true steady state value. For higher $r$ values, a clear metastable plateau is observed, with the same plateau value across different initial conditions. Notably, this metastable state occurs at $\langle w \rangle_c \approx 3N/4 = 75$, which is precisely the steady state value of the ideal OTOC ($r=1$) case. However, trajectories starting from lower initial conditions may not reach this metastable plateau before converging to the final steady state. The $\kappa$ dependence is omitted as it merely rescales time as discussed in the main text.}
    \label{fig:metastable_full}
\end{figure}

\begin{figure}
    \centering
    \begin{tabular}{cc}
        \includegraphics[width=0.48\textwidth]{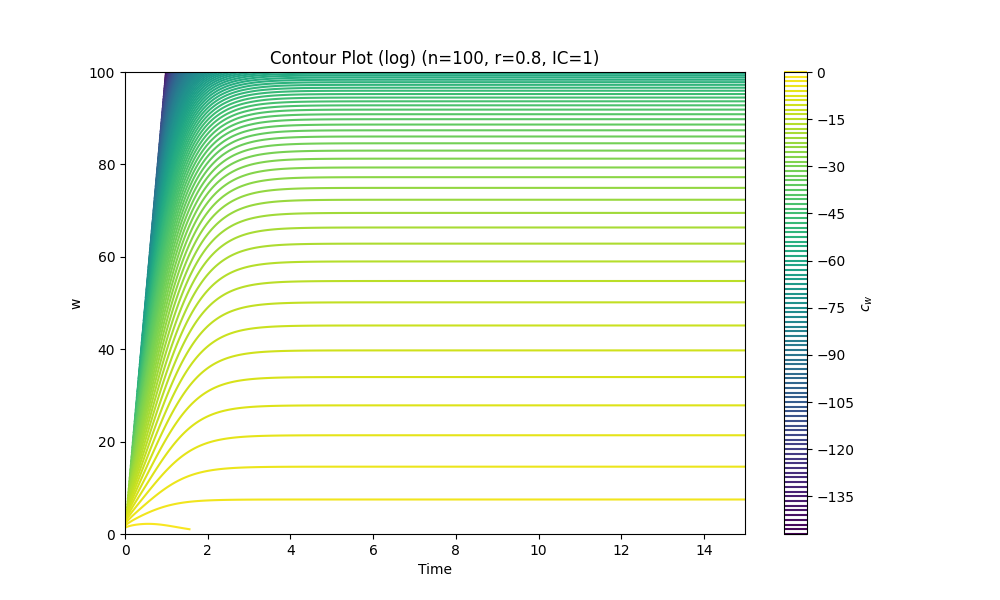} &
        \includegraphics[width=0.48\textwidth]{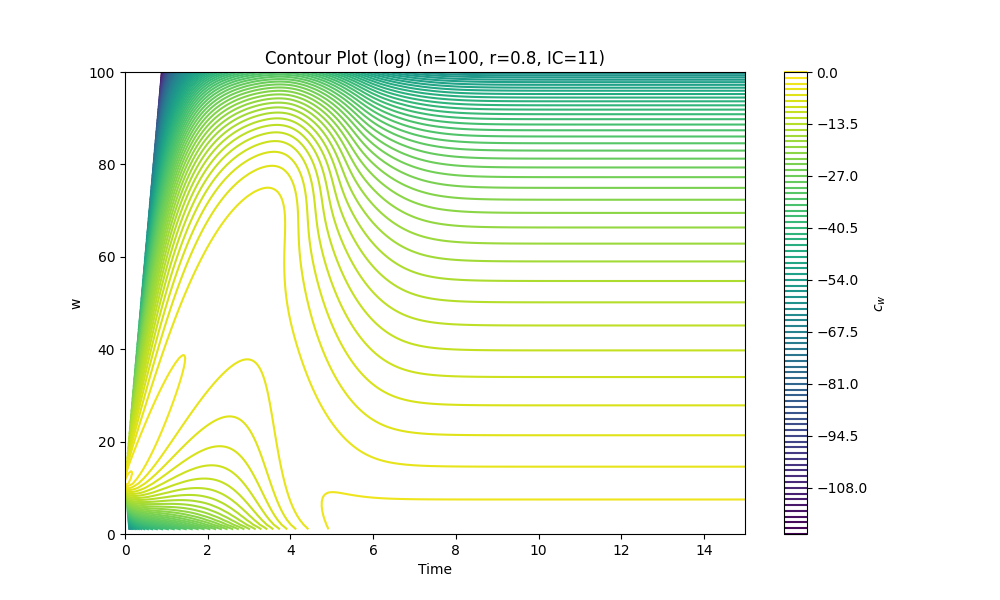} \\
        (a) & (b) \\
        \includegraphics[width=0.48\textwidth]{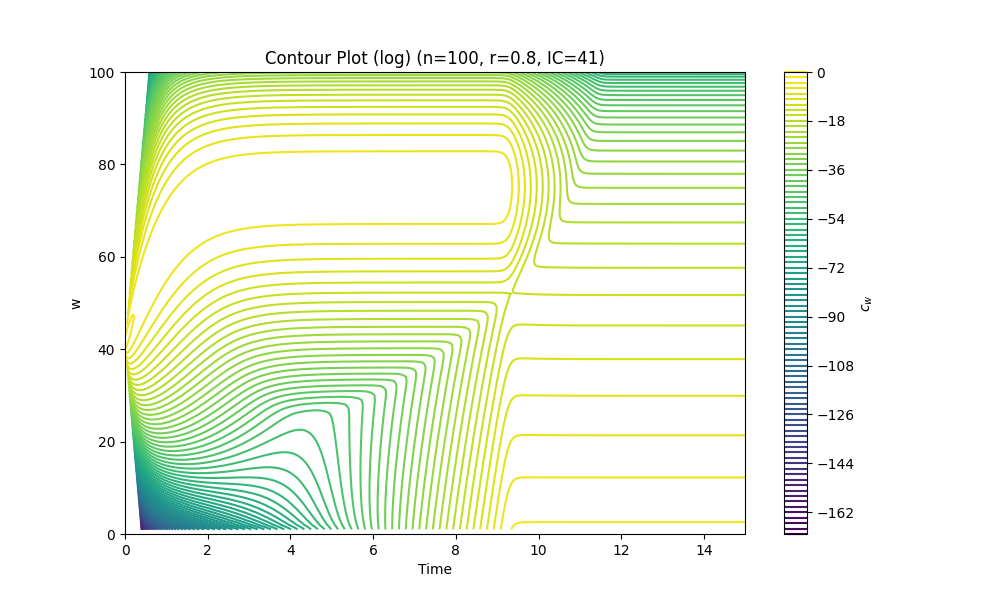} &
        \includegraphics[width=0.48\textwidth]{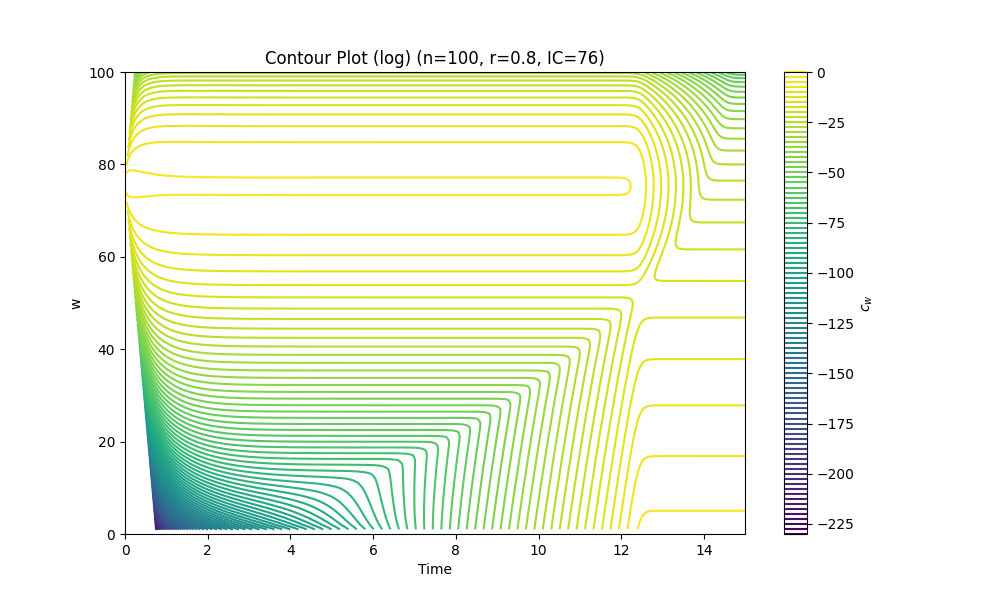} \\
        (c) & (d)
    \end{tabular}
    \caption{Evolution of probability distributions for different weights plotted as contour plots in logarithmic scale over 15 time units. The y-axis represents weight and the x-axis represents time for a system with $N=100$ qubits and correlation $r=0.8$. Four different initial conditions are shown: (a) IC=1, (b) IC=11, (c) IC=41, and (d) IC=76. In panel (a) with IC=1, no metastable state is observed. In panel (b) with IC=11, the probability mass never reaches the metastable state before the redistribution term (renormalization by echo) takes over. In panels (c) and (d) with IC=41 and IC=76 respectively, we observe regions of exponential probability decay/increase characterized by evenly spaced vertical lines, indicating rapid redistribution of probability.}
    \label{fig:average_weight_full}
\end{figure}

\begin{figure}
    \centering
    \includegraphics[width=\textwidth]{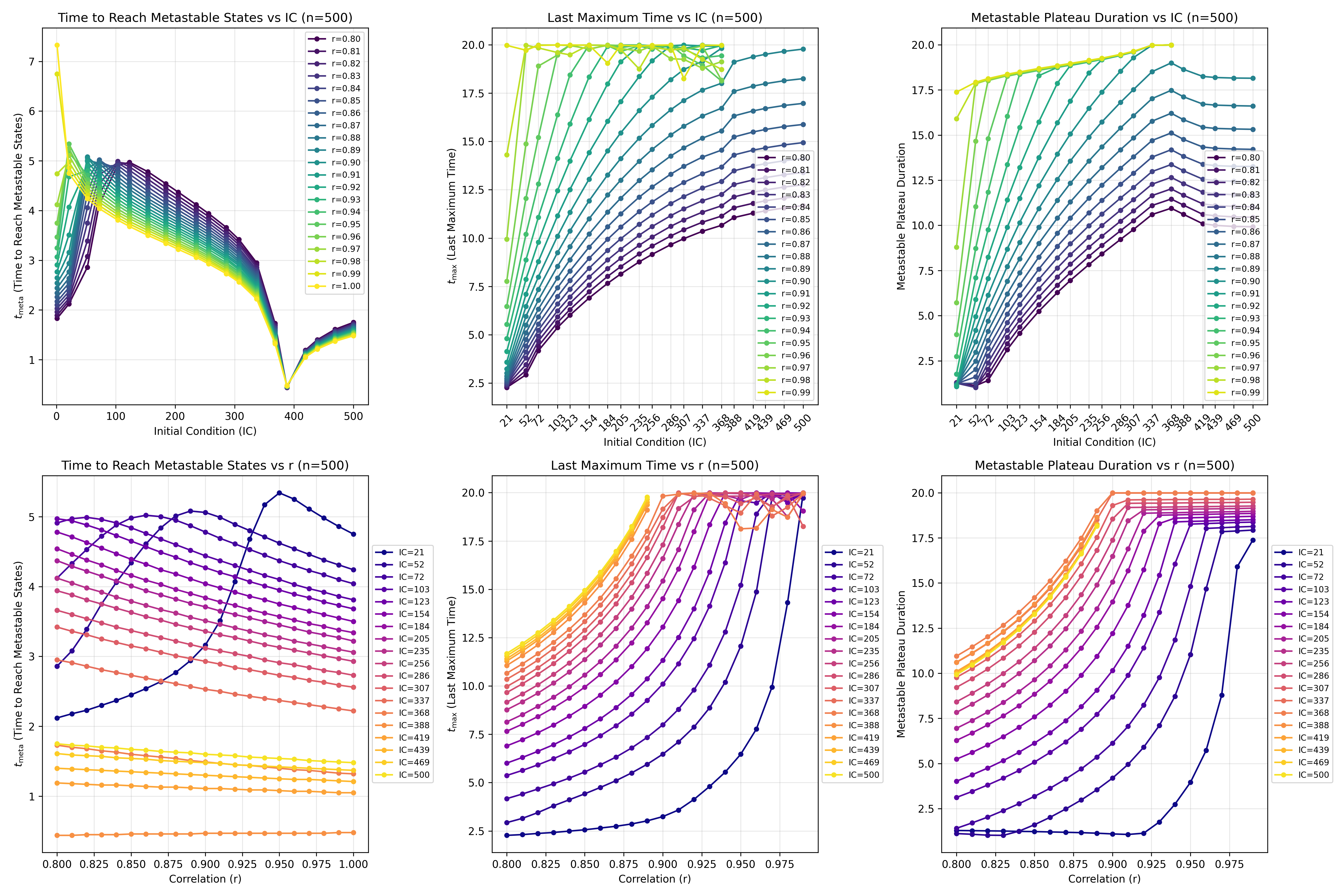}
    \caption{Analysis of metastable state dynamics for a system with $N=500$ qubits. The first column shows the time to reach the metastable plateau, the second column shows the time when the system leaves the plateau, and the third column shows the duration of the metastable plateau (difference between the first two times). When no prolonged metastable plateau exists, the time to reach the plateau is defined as the time to reach maximum $\expval{w}_c$ values. Note that minimum durations in the third column are not exactly zero due to a small tolerance value in the numerics used to determine plateau entry/exit. The top row shows these times as functions of initial condition (IC), while the bottom row shows them as functions of correlation $r$. The time to reach the metastable plateau is primarily determined by operator growth dynamics (OTOC behavior), while the time to leave the plateau is mainly governed by the redistribution effect from the echo.}
    \label{fig:metastable_duration_combined}
\end{figure}

In this appendix, we provide supplementary numerical results for the full dynamics of the master equation. The dilute limit allows for analytical solutions as presented in this work, but an analytical solution for the full dynamics remains unclear. These numerical results serve to complement our analytical understanding by capturing the behavior beyond the dilute approximation, and for potential future investigation. 

First, an example calculation for $N=200$ ROTOC is plotted in Fig.~\ref{fig:metastable_full}. We note that in the dilute limit ($w_0\ll N$), the ROTOC behaviour is the same as those considered in the main text (see Fig.~\ref{fig:metastable}). We observe that , in the case when the initial weight leaves the dilute limit, the metastable plateau is extended and maintains its value at $3N/4$, which is identical to that of the ideal OTOC. From subfigures (b) and (c), when the initial weight is in the intermediate regime $1 \ll w_0 \ll N$, the metastable lifetime shows a linear dependence on the initial weight $w_0$.

In Fig.~\ref{fig:average_weight_full}, we show examples of the contour plots of probability mass at all weights over time from the same set of parameters considered above. We observe distinct behaviors across different parameter regimes. In the $r=1.0$ ideal OTOC case, the redistribution effect from echo never activates, so the dynamics moves monotonically to the steady state and remains there.

In subfigures (c) and (d), with a high enough initial weight, we observe distinct dynamical regimes. Examining the full equation of motion for the renormalized probability distribution $c_w$ Eq.~\eqref{eq:c_eom},
\begin{equation}
    \dv{c}{t} = Mc+\mu c
\end{equation}
the redistribution of probability is controlled by the term $\sim\langle w \rangle_c \cdot c_w$. When the target weight has zero probability, redistribution will not occur.
    
When we start with high initial weight, the process of probability mass flowing downward to smaller weight is $1/N$ suppressed, meaning at higher initial weight, the redistribution effect will activate later. Between the metastable state and the true steady state in the later portion of subfigures (c) and (d), we observe the exponential decay in probability at all weights caused by the exponential decay in echo.

In Fig.~\ref{fig:metastable_duration_combined}, we analyze the duration of this metastable plateau for the full dynamics, defined as the difference between the time to reach and the time to leave the plateau. This differs slightly from the dilute limit case considered in the main text (Sec.~\ref{subsec:results_metastable}), where we considered the lifetime to be simply the time to leave the metastable state. We note that the time to reach the metastable plateau is largely determined by the effects from operator spreading, as the redistribution term is, and the time to leave the plateau is largely determined by the redistribution effect contributed by the echo. This can be equivalently interpreted as the decay in fidelity over time.
    
The $\sim-\log w_0$ dependence discussed in the main text Eq.~\eqref{eq:metastable_lifetime} is barely observed with smaller $r$ values, as the system size we consider here is small ($N=100$).
This equivalence of limits is described by the finite size crossover (see Appendix~\ref{app:dilute_ftc}), where $r<r_*$ exhibits stronger dilute limit traits as this is the regime when curves are still growing, while $r>r_*$ represents curves that have saturated ,see Fig.~\ref{fig:r_dilute_equivalence}.

\end{widetext}

\end{document}